\documentclass[12pt]{article}

\usepackage{amsmath}
\usepackage{amsthm}
\usepackage{amsfonts}
\usepackage{mathtools}
\usepackage[english]{babel}
\usepackage{footnote}
\makesavenoteenv{tabular}
\usepackage{floatpag}
\usepackage{graphicx,psfrag,epsf}
\usepackage{enumerate}
\usepackage[noend]{algpseudocode}
\usepackage{natbib}
\usepackage{url} 
\usepackage{comment}
\usepackage{booktabs}   
\usepackage{bm}
\newcommand{\rmnum}[1]{\romannumeral #1}
\usepackage{algorithmicx,algorithm}
\usepackage{multirow}
\usepackage{threeparttable}
\usepackage{wasysym}
\usepackage{framed}
\usepackage{pgfgantt,rotating}
\usepackage{multirow}
\usepackage{blindtext}
\usepackage{times}

\usepackage{epsf,subfigure,pstricks,pst-node}
\usepackage{float}
\usepackage[outdir=./]{epstopdf}
\usepackage{verbatim} 

\newtheorem{theorem}{Theorem}
\newtheorem{lemma}{Lemma} 
\newtheorem{proposition}{Proposition} 
\newtheorem{remark}{Remark}

\newtheorem{definition}{Definition}

\newtheorem{assumption}{Assumption}

\newcommand{\ind}{\perp \!\!\! \perp }
\newcommand{\nind}{\not\!\perp\!\!\!\perp}
\newcommand{\mC}{\mathcal{C}}
\newcommand{\mG}{\mathcal{G}}
\newcommand{\mP}{\mathcal{P}}
\newcommand{\mI}{\mathcal{I}}
\newcommand{\mN}{\mathcal{N}}
\newcommand{\mS}{\mathcal{S}}

\def\T{{ \mathrm{\scriptscriptstyle T} }}

\def\K{\mathcal{K}}
\newcommand{\bigCI}{\mathrel{\text{\scalebox{1.07}{$\perp\mkern-10mu\perp$}}}}

\usepackage{amsmath}
\usepackage{graphicx}
\usepackage{enumerate}
\usepackage{natbib}
\usepackage{url} 

\newcommand{\blind}{1}

 \date{}

\begin{document}

\def\spacingset#1{\renewcommand{\baselinestretch}%
{#1}\small\normalsize} \spacingset{1}


\if1\blind
{
  \title{\bf Ultra-high Dimensional Variable Selection for Doubly Robust Causal Inference}
  \author{ }
  \maketitle
\begin{center}
  \author{\large Dingke Tang \\
  \vspace{10pt}
  Department of Statistical Sciences, University of Toronto}\\ \vspace{10pt}
    \author{\large Dehan Kong \\
    \vspace{10pt}
  Department of Statistical Sciences, University of Toronto}\\ \vspace{10pt}
    \author{\large Wenliang Pan \\
    \vspace{10pt}
  Department of Statistical Science, School of Mathematics, Sun Yat-Sen University}\\ \vspace{10pt}
  \author{\large Linbo Wang \\
  \vspace{10pt}
Department of Statistical Sciences, University of Toronto}\\
\end{center}
\newpage
} \fi

\if0\blind
{
  \bigskip
  \bigskip
  \bigskip
  \begin{center}
    {\LARGE\bf Outcome model free causal inference with ultra-high dimensional covariates}
\end{center}
  \medskip
} \fi

\bigskip
\spacingset{1.5} 
\begin{abstract}
Causal inference has been increasingly reliant on observational studies with rich covariate information. To build tractable causal procedures, such as the doubly robust estimators, it is imperative to first extract important features from high or even ultra-high dimensional data. In this paper, we propose causal ball screening for confounder selection from modern ultra-high dimensional data sets. Unlike the familiar task of variable selection for prediction modeling, our confounder selection procedure aims to control for confounding while improving efficiency in the resulting causal effect estimate. Previous empirical and theoretical studies suggest excluding causes of the treatment that are not confounders. Motivated by these results, our goal is to keep all the predictors of the outcome in both the propensity score and outcome regression models. A distinctive feature of our proposal is that we use an outcome model-free procedure for propensity score model selection, thereby maintaining double robustness in the resulting causal effect estimator. Our theoretical analyses show that the proposed procedure enjoys a number of  properties, including model selection consistency and point-wise normality. Synthetic and real data analysis show that our proposal performs favorably with existing methods in a range of realistic settings. Data used in preparation of this article were obtained from the Alzheimer’s Disease Neuroimaging Initiative (ADNI) database.
\end{abstract}

%

\noindent%
\vfill

\newpage
\spacingset{1.5} 
\section{Introduction}
Modern observational databases  hold great promise for drawing causal conclusions. In these
studies, both the treatment and outcome of interest are often associated with some baseline covariates, called confounders. Insufficient adjustment for confounders leads to biased causal effect estimates. In their seminal work, Rosenbaum and Rubin (1983) showed that the propensity score, defined as the probability of assignment to a particular treatment conditional on baseline covariates, can be used to remove bias due to observed confounders. {\cite{robins1994estimation} proposed a doubly robust method that combines outcome regression and propensity score modeling. Their estimator has been shown to enjoy favorable theoretical properties under correct specification of the outcome regression and/or propensity score models.}

Traditionally, specifications of the propensity score  {and outcome regression} models are typically driven by expert knowledge. However, this is becoming increasingly difficult in modern applications, where researchers are often presented with high or even ultra-high dimensional covariates. For example, a popular database for studying potential risk factors of Alzheimer's disease is available through the Alzheimer’s Disease Neuroimaging Initiative (ADNI). The ADNI study collects rich covariate information such as clinical and behavioral covariates, and genetic information including millions of SNPs. The dimension of covariates $p$ is much larger than the sample size $n$. This is known as ultra-high dimensional data in the literature, as classical variable selection methods for high-dimensional data such as the Lasso are not feasible due to computational complexity.

In response to these challenges,  there has been a growing interest in developing data-driven procedures for covariate selection in causal inference. A central aim of these methods is to reduce bias and improve efficiency in the final causal effect estimator \citep{witte2019covariate}. This is in sharp contrast to covariate selection in prediction modeling \citep[e.g.][]{tibshirani1996,fan2008sure}, where the goal is  to find a sparse representation of the association structure with good prediction accuracy. In particular, a good prediction model for the propensity score includes all strong predictors of the treatment. However, theoretical results and empirical evidence imply that inclusion of variables associated with treatment but not the outcome may inflate the variance of the resulting causal effect estimates 
\citep[e.g.][]{brookhart2006variable,de2011covariate,schnitzer2016variable,rotnitzky2020efficient}. Such variables are commonly referred to as instrumental variables (IVs). In particular, \cite{hahn1998role,hahn2004functional} showed that the semiparametric efficiency bound for estimating the average causal effect may be reduced if some covariates are known to be instrumental variables.  Although  selection of instrumental variables may lead to non-uniform inference \citep[e.g.][]{leeb2005model,moosavi2021costs},  it has been reported that in many practical settings, 
selection of instrumental variables improves the estimate in a mean squared error sense \citep{brookhart2006variable}; see the end of Section \ref{theories} for a detailed discussion.


Motivated by these results, various procedures have been developed for selecting proper variables into the propensity score and outcome regression models. A naive approach is prediction modeling for the outcome \citep[e.g.][]{tibshirani1996} while specifying the treatment as a fixed covariate in the model. When used with a small sample size, it may miss confounders that are  weakly associated with the outcome but strongly associated  with the treatment \citep{wilson2014confounder}. The omission of such variables leads to  bias but reduces standard error; in some scenarios, it even reduces the mean squared error \citep{brookhart2006variable}. Alternatively, \cite{zigler2014uncertainty} proposed a Bayesian model averaging approach based on a PS model and an outcome model conditional on the estimated PS and baseline covariates. \cite{shortreed2017outcome} proposed the outcome-adaptive Lasso, which penalizes  the coefficients of a propensity score model inversely proportional to their coefficients in a separate outcome regression model. \citet{ertefaie2018variable} proposed a variable selection method using a penalized objective function based on both a linear outcome and a logistic propensity score model. Under a sparse linear outcome model, \cite{antonelli2019high} proposed a Bayesian approach that uses continuous spike-and-slab priors on the regression coefficients corresponding to the confounders. The validity of  these confounder selection methods relies on  correct specification of the outcome regression model.
If the same  outcome model was used for both PS model selection and causal effect estimation, then the resulting ``doubly robust'' estimator \citep{robins1994estimation} is no longer doubly robust.  In particular, if the outcome regression model is incorrect, then the selected PS model may miss important confounders, so that estimates from the PS model, outcome regression model, and hence the ``doubly robust'' estimator may all be biased. 
An additional pitfall of existing methods is that none of them is well-suited for covariate selection from an ultra-high dimensional feature set, such as the one collected by the ADNI study. For ultra-high dimensional data, penalization or Bayesian selection methods face challenges in computational cost and estimation accuracy \citep{fan2008sure}.

In this paper, we  propose Causal Ball Screening (CBS), a novel doubly robust causal effect estimating procedure that combines an outcome model-free screening step motivated by the ball covariance \citep{pan2018generic,pan2019ball} with a refined selection and doubly robust estimation step. In contrast to aforementioned approaches that aim to exclude instruments, our proposal for propensity score model selection is outcome model-free:  it does not require specifications of the outcome regression model, nor does it involve any smoothness assumptions on the outcome regression. As a result, the resulting causal effect estimator is doubly robust. 
 Furthermore, to the best of our knowledge, our method is the first in the causal inference literature that applies to ultra-high dimensional settings. 

The rest of the article is organized as follows. Section \ref{preliminary} introduces background on the target adjustment set in doubly robust causal effect estimation, and  the ball covariance.  In Section \ref{method}, we 
introduce our CBS procedure for doubly robust causal effect estimation with ultra-high dimensional covariates. Section \ref{theories} provides theoretical justifications for the CBS. Simulation studies in Section \ref{sim}  compare our proposal with several state-of-art methods in their finite-sample performance. In Section \ref{realdata}, we apply our method to the ADNI study and estimate the causal effect of tau protein level in cerebrospinal fluid on Alzheimer's behavioral score while accounting for ultra-high dimensional covariates. 
We end with a brief discussion in Section \ref{sec:discussion}.

\section{Background}\label{preliminary}

\subsection{The propensity score}

Following the potential outcome framework, we use $ D $ to denote a  binary treatment assignment, $X = (X^{(1)},\ldots,X^{(p)})$ to denote baseline covariates, and $ Y(d) $ to denote the outcome that would have been observed under treatment assignment $d$ for $d=0, 1$. We assume that the covariates $X$  are ultra-high dimensional in the following sense.

\begin{definition}[Ultra-high dimensionality]
We say covariates $X$ are ultra-high dimensional if the number of covariates $p = O\left\{\exp(n^\iota)\right\}$ for some constant $\iota>0$, where $n$ is the sample size.

\end{definition}

We make the stable unit treatment value assumption \citep{rubin1980comment}.

\begin{assumption}[Stable unit treatment value assumption] The potential outcomes for any unit
do not vary with the treatments assigned to other units; for each unit, there are no different forms
or versions of each treatment level, which leads to different potential outcomes.
\label{assumption:sutva}
\end{assumption}

Under Assumption \ref{assumption:sutva}, the observed outcome $Y$ satisfies $Y = DY(1) +(1-D)Y(0)$ \citep{vanderweele2013causal}. 
Suppose we observe $n$ independent samples from the joint distribution of $(X, D, Y)$, denoted by $(X_i, D_i, Y_i), i=1,\ldots, n.$ 
We are interested in estimating the average causal effect (ACE) $\Delta = E\{Y(1) - Y(0)\}$. The ACE can be non-parametrically identified under the following assumptions.
\begin{assumption}[Weak Ignorability] \label{assumption:ignorability}
 There exists $ X^\mS $ such that $D \bigCI Y(d) \mid X^\mS$ for $d = 0, 1$, where $S \subset\{1,\ldots, p\}.$
\end{assumption}
\begin{assumption}[Positivity]
\label{assumption:positivity}
$0 < c \leq P(D = 1 \mid X^\mS) \leq 1- c < 1$, where $c\in (0,1)$.
\end{assumption}

\cite{rosenbaum1983central} introduced the notion of propensity score $e(X^\mS) = P(D=1\mid X^\mS)$  and showed that under Assumptions \ref{assumption:ignorability} and \ref{assumption:positivity}, adjusting for the propensity score is sufficient to remove confounding:
$
        D\ind Y(d) \mid e(X^\mS), d=0,1.
$



\subsection{Doubly robust estimation} \label{sec:psweight}
 Let $\widehat{e}_i = \widehat{e}(X_i^\mS)$ be the estimated propensity score, and $\widehat{b}_d(X_i^\mS)$ be the estimate of outcome regression $E(Y \mid D=d,X_i^\mS)$. 
  The classical doubly robust estimator \citep{robins1994estimation} is defined as
\begin{equation}
\label{eqn:dr}
\widehat{\Delta} = \dfrac{1}{n} \sum\limits_{i=1}^n \dfrac{D_i Y_i - (D_i-\widehat{e}_i) \widehat{b}_1(X_i^\mS)}{\widehat{e}_i} - \dfrac{1}{n} \sum\limits_{i=1}^n \dfrac{(1-D_i) Y_i + (D_i-\widehat{e}_i)\widehat{b}_0(X_i^\mS)}{1-\widehat{e}_i}.
\end{equation}
It has been shown that $\widehat{\Delta}$ is locally semiparametric efficient in the sense that if both the propensity score and outcome regression model estimates converge to their corresponding true values at sufficiently fast rates, then asymptotically the variance of $\widehat{\Delta}$ achieves the semiparametric efficiency bound \citep[e.g.][]{chernozhukov2018double}. Furthermore, it is consistent if either the outcome regression or the propensity model is correctly specified.

\subsection{Target adjustment set}
\label{Target set}
We now discuss the target adjustment set of variables to include in the propensity score and outcome regression models, in order to eliminate bias and reduce variance in the resulting doubly robust causal effect estimates.  We first divide the covariates into four disjoint subsets under the framework of a causal directed acyclic graph (DAG) \citep{pearl2009causality}. Relevant background on the DAG framework is provided in Section \ref{sec:Details in target adjustment set} of the Supplementary Material.  

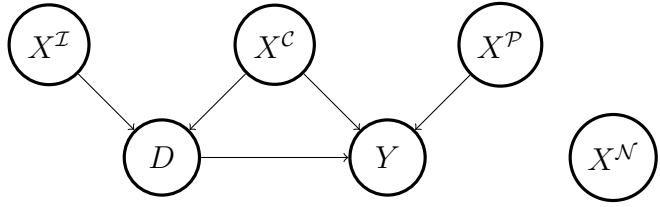
\begin{figure}
    \centering
 \begin{tikzpicture}[
roundnode/.style={circle, draw=black, very thick, minimum size=10mm},
]
\node[roundnode]        at (0, 0)  (D)  {$D$};
\node[roundnode]        at (3, 0)  (Y) {$Y$};
\node[roundnode]        at (6, 0)  (XN)  {$X^\mN$};
\node[roundnode]        at (-1.5, 1.5) (XI)  {$X^\mI$};
\node[roundnode]        at (1.5, 1.5)  (XC)  {$X^\mC$};
\node[roundnode]        at (4.5, 1.5)  (XP)  {$X^\mP$};
\draw[->] (D) -- (Y);
\draw[->] (XI) -- (D);
\draw[->] (XC) -- (D);
\draw[->] (XC) -- (Y);
\draw[->] (XP) -- (Y);

\end{tikzpicture}\caption{A causal directed acyclic graph  illustrating the four types of baseline covariates: confounders $X^{\mathcal{C}}$, precision variables $X^{\mathcal{P}}$, instrumental variables $X^{\mathcal{I}}$ and null variables $X^\mN$}
    \label{DAG}
\end{figure}

Let $X^\mC = \{X^{(j)} \in pa(Y): D  \text{ and }  X^{(j)}   \text{ are  d-connected  given }  pa(Y)\setminus \{X^{(j)}\} \}$, where $pa(Y)$ denotes the parents of $Y$.  In the following we shall refer to $X^\mC$ as confounders, $X^\mP = pa(Y) \setminus X^\mC$ as precision variables, $X^\mI = pa(D) \setminus X^\mC$ as instrumental variables, and $X^\mN = X \setminus (X^\mC \cup X^\mP \cup X^\mI)$ as null variables. 
Figure \ref{DAG} provides the simplest causal diagram associated with these definitions. 

\begin{remark} \label{ref: luna}
Under Assumptions \ref{assumption:sufficiency} and \ref{assumption:order}, our definition of confounders is a special case of the reduced covariate set labelled as 
$Z$ in \cite{de2011covariate}.  The instrumental variable is commonly used to estimate causal effects when Assumption \ref{assumption:ignorability} may be violated. Under the causal sufficiency assumption in the Supplementary Material, our definition of instrumental variable here coincides with that in the literature \citep[e.g.][]{wang2018bounded}. 
\end{remark}

If one already adjusts for all the confounders, then  conditioning on additional precision variables and instrumental variables still renders the treatment $D$ and potential outcome $Y(d)$ conditional independent; see  Proposition \ref{prop:ignorability}  in the Supplementary Material and \citet[][Proposition 3]{de2011covariate} for detailed arguments. Adjusting for the null variables may, however, introduce  bias. Consider the causal DAG with $D\rightarrow Y$ and  $D\leftarrow X^\mI \rightarrow X^{collider} \leftarrow X^\mP \rightarrow Y.$ Under our definitions, $X^{collider}$ is a null variable and the empty set is a valid adjustment set. 
But adjusting for $X^{collider}$ may lead to a biased causal effect estimate due to collider bias \citep{pearl2009causality}.

    


Previous theoretical and empirical findings suggest that  inclusion of instrumental variables in addition to confounding variables in the adjustment set may result in efficiency loss \citep{hahn1998role,hahn2004functional, brookhart2006variable,de2011covariate}. 


\begin{proposition}\citep{hahn2004functional}\label{prop:DR bound}  Let Assumptions \ref{assumption:sutva}--\ref{assumption:positivity} and Assumptions \ref{assumption:sufficiency}--\ref{assumption:order} in the Supplementary Material hold, and suppose we have two restrictions (R.1) $\mP \neq \emptyset$ and (R.2) $\mI\neq \emptyset$. Then the semiparametric efficiency bound for estimating $\Delta$ when (R.1) holds is equal to the bound without the restriction, and the semiparametric efficiency bound for estimating $\Delta$ when (R.2) holds is lower than the bound without the restriction. 

\end{proposition}


Following these results, our target adjustment set for propensity score and outcome modeling is $\mathcal{A} = \mC \cup \mP$. We include the confounders  in $\mathcal{A}$ to avoid confounding bias, and exclude null variables to avoid potential collider bias. Motivated by 
Proposition \ref{prop:DR bound}, we shall 
exclude instruments in $\mathcal{A}$ to reduce variance in the resulting doubly robust causal effect estimator. As inclusion of any subset of  precision variables $\mathcal{P}$ in the adjustment set does not introduce bias or affect efficiency, we do not attempt to exclude precision variables in the adjustment set.

\begin{remark}
\label{remark:pv}
  In the case that the precision variables are not sparse in the covariate set (but the confounders are),  one would need to further exclude precision variables in the adjustment set. 
\end{remark}

\subsection{The ball covariance}

The ball covariance is a generic measure of dependence in Banach space with many desirable properties \citep{pan2019ball}. Importantly, it is entirely model-free for data in Euclidean spaces, and its empirical version is easy to compute as a test statistic of independence. Furthermore, compared with other measures of dependence between two random variables such as the mutual information \citep{cover2012elements} and distance correlation \citep{szekely2007measuring}, the ball correlation does not require the random variables to have finite moments. It hence provides robustness for data with a heavy-tailed distribution \citep{pan2018generic}.

Specifically, let $X$,$Y$ be two random variables on separable Banach spaces $(\mathcal{X},\rho)$ and $(\mathcal{Y},\xi)$, respectively, where $\rho$ and $\xi$ are distance functions in the respective spaces. And let $\theta$, $\mu$, $\nu$ be probability measures induced by $(X,Y)$, $X$, $Y$, respectively. Denote $\overline{B}_\rho(x_1,x_2)$ a closed ball in space $(\mathcal{X},\rho)$ centering in $x_1$ with radius $\rho(x_1,x_2)$, and $\overline{B}_\xi(y_1,y_2)$ a closed ball in space $(\mathcal{Y},\xi)$ centering in $ y_1$ with radius $\xi(y_1,y_2)$. 

\begin{definition}The ball covariance is defined as the square root of $BCov^2(X,Y)$, which is an integral of
the Hoeffding’s dependence measure on the coordinate of radius
over poles:
\begin{equation*}
\begin{split}
BCov^2(X,Y) & = \int(\theta-\mu\otimes\nu)^2\{\overline{B}_\rho(x_1,x_2)\times \overline{B}_\xi(y_1,y_2)\} \theta(dx_1,dy_1)\theta(dx_2,dy_2),
\end{split}
\end{equation*}
\end{definition}
where $\mu\otimes\nu$ is a product measure on $\mathcal{X}\times\mathcal{Y}$. 

 Let $\delta_{ij,k}^X$ = $I\{X_k\in\overline{B}_\rho(X_i,X_j)\}$, where $ I(\cdot)$ is an indicator function. Further define $\delta^X_{ij,kl} = \delta^X_{ij,k}\delta^X_{ij,l}$ and $\xi^X_{ij,klst} = (\delta^X_{ij,kl}+\delta^X_{ij,st}-\delta^X_{ij,ks}-\delta^X_{ij,lt})/2$. We can similarly define $\delta^Y_{ij,k},\delta^Y_{ij,kl},\xi^Y_{ij,klst}$. 

\begin{proposition} \label{prop1}\citep[Separability property,][]{pan2019ball} Let ($X_i$, $Y_i$), i = 1, 2, ..., 6 be i.i.d samples from the joint distribution of ($X,Y$). Then
$
BCov^2(X,Y) = E(\xi^X_{12,3456}\xi^Y_{12,3456}).
$
\end{proposition}



\cite{pan2019ball} show that  $BCov(X,Y)=0$ if and only if $X$ is independent of $Y$. Therefore, the ball covariance can be used to perform the independent test. 



We next introduce the empirical version of ball covariance. 

\begin{definition} (Empirical ball covariance) The empirical ball covariance $BCov_{n}$($X$, $Y$) is defined as the square root of: $
BCov^2_n(X,Y)=\dfrac{1}{n^6}\sum^n_{i,j,k,l,s,t=1}\xi^X_{ij,klst}\xi^Y_{ij,klst}.
$
\end{definition}



\section{Causal ball screening}\label{method}

 In this section, we develop causal ball screening (CBS), a two-step procedure for covariate selection and causal effect estimation. The first step involves a generic sure independence screening procedure to screen out most null and instrumental variables while keeping all the confounding and precision variables. 
 This procedure is based on the conditional ball covariance, a novel concept we introduce based on the ball covariance \citep{pan2018generic,pan2019ball}. The second step is a refined selection and estimation step that  excludes the null and instrument variables  and estimates the average causal effect.

\subsection{Conditional ball covariance screening}\label{sec:condball}

When the candidate feature set is ultra-high dimensional, a common strategy is to use the sure independence screening procedure based on marginal \citep{fan2008sure} or conditional correlations \citep{barut2016conditional}. From Figure \ref{DAG}, if the DAG is faithful, then one can read off the following (conditional) (in)dependences:
\begin{flalign}
    \label{eqn:ind} X^\mC \nind Y,\qquad \ \  X^\mP \nind Y, \quad \ \ & X^\mI \nind Y, \quad \ \ X^\mN \ind Y;  \quad \ \ \\
    \label{eqn:cind} X^\mC \nind Y\mid D, \ \  X^\mP \nind Y\mid D, \ \ & X^\mI \nind Y\mid D, \ \  X^\mN \ind Y\mid D. \ \
\end{flalign}

On the surface, it seems that independence screening based on  \eqref{eqn:ind} or \eqref{eqn:cind} works equally well. In practice, however, note that the dependence between $X^\mI$ and $Y$ after conditioning on $D$ is induced by collider  bias. Previous qualitative analyses \citep{ding2015adjust} and numerical analysis \citep{liu2012implications} show that collider bias tends to be small in many realistic settings. Consequently, we perform our screening based on \eqref{eqn:cind} under the assumption that the instrumental variables $X^\mI$ have weaker dependence with the outcome $Y$ after conditioning on the treatment variable $D$, and hence are more likely to be screened out by the conditional independence screening.

To perform conditional independence screening based on \eqref{eqn:cind}, we first introduce the notion of conditional ball covariance. 
Let $\omega = P(D=1)$ be the probability of receiving treatment. Let $X^{(d)}, Y^{(d)}, d=0,1$ be random variables such that $
         (X^{(d)}, Y^{(d)}) \stackrel{d}{=} (X,Y\mid D=d), d=0,1.$

The conditional ball covariance between $X$ and $Y$ given $D$ is defined as the square root of
\begin{equation*}
\label{eqn:cond_bcov}
\begin{split}
    &BCov^2(X,Y\mid D) = \omega BCov^2(X^{(1)},Y^{(1)}) + (1-\omega) BCov^2(X^{(0)},Y^{(0)}).
\end{split}
\end{equation*}
Analogously, we can define the sample version of the conditional ball covariance. Let $n_1 = \sum_{i=1}^n D_i$ be the number of subjects who receive treatment and $n_0 = n-n_1$. Let $\widehat{\omega}=n_1/n$ be the empirical estimator of $\omega$.

\begin{definition} The empirical conditional ball covariance $BCov_{n}(X, Y\mid D)$ is defined as the square root of:
$BCov^2_n(X,Y\mid D) = {\widehat{\omega}}\sum_{(i,j,k,l,s,t):D_i,D_j,D_k,D_l,D_s,D_t=1} \xi^X_{ij,klst}\xi^Y_{ij,klst}/ {n_1^6}+ 
(1-\widehat{\omega})\sum_{(i,j,k,l,s,t):D_i,D_j,D_k,D_l,D_s,D_t=0} \xi^X_{ij,klst}\xi^Y_{ij,klst}/{n_0^6}.$
\end{definition}

The following proposition  is an extension of Lemma 2.1 in \citet{pan2018generic}. 
\begin{proposition}\label{prop:independence}
\label{identification prop}
  $BCov(X,Y\mid D) = 0 \Leftrightarrow X\ind Y \mid D.$
\end{proposition}

To perform conditional ball covariance screening, as summarized in the first two steps of Algorithm \ref{alg:cbs}, we first calculate the empirical conditional ball covariance between the outcome $Y$ and each baseline covariate $X^{(j)}$, $j=1,\ldots, p,$ and then select $q$ baseline covariates with the largest ball covariance into the next step. 
 Let $\K$ be the selected set after the screening step.
 Without loss of generality, we assume  $\K = \{1,2,\ldots,q\}$.

\subsection{Refined selection and doubly robust estimation}
\label{variable selection}

There may   be instrumental variables and null variables remaining in the set $\K$ after the screening step. 
We now propose a second refined selection step to further exclude these variables. To estimate parameters in the outcome regression models on $b_d(X),\; d=0,1$, we  use the Lasso estimator \citep{tibshirani1996} that 
 \begin{equation}\label{eqn:Lasso}
     \widehat{\alpha}^{(d)}_{\mathcal{K}} = \mathop{\rm argmin}\limits_{ {\alpha}_\K}\left\{\sum_{i: D_i=d}(Y_i-X_{i,\mathcal{K}}^\T\alpha_{\mathcal{K}})^2+\lambda_Y^{(d)}||\alpha_\mathcal{K}||_1\right\}, d=0,1.
 \end{equation}
 In practice, we use $10$-fold cross-validation to select the tuning parameters $\lambda_Y^{(d)}, d=0,1$.

Refined selection for the propensity score model is more involved. As we explained in the introduction, for double robustness of the resulting causal effect estimator, selection of the PS model should not depend on the outcome regression model.
So we shall apply the idea of  adaptive Lasso \citep{zou2006adaptive,shortreed2017outcome} to our setting. Specifically, let
\begin{equation}
\label{eqn:selection}
\begin{aligned}
    \widehat{ {\beta}}_\K&= \mathop{\rm argmin}\limits_{ {\beta}_\K}\left(  \sum_{i=1}^n\left[D_i\log\left\{\dfrac{1-e(X_{i,\K};{ {\beta}}_\K)}{e(X_{i,\K};{ {\beta}}_\K)}\right\} - \log\left\{1-e(X_{i,\K};{ {\beta}}_\K)\right\}\right]\right. \left.+\lambda_D\sum_{j=1}^{q}\dfrac {1}{\widehat{\omega}_j}|\beta_j|\right),
\end{aligned}
\end{equation}
where $\widehat{\omega}_j$ is a nonparametric estimator of the importance of covariate $X^{(j)}$ in the outcome model, $\lambda_D$ is a tuning parameter and the propensity score follows a logistic regression model that  $e(X;\beta) = \text{expit}(X^T \beta) = \exp(X^T\beta)/\{1+\exp(X^T \beta)\}$. In practice, $\widehat{\omega}_j$ can be obtained based on the inverse of the conditional mutual information \citep{berrett2019efficient}, the conditional distance correlation \citep{wang2015conditional}, or the conditional ball covariance introduced in Section \ref{sec:condball}. In the simulations and data analysis, we let $\widehat{w}_j={ |\widehat{z}\;BCov_n^2(X^{(j)},Y\mid D)|^\gamma}$, where $\gamma$ is a tunning parameter
and $\widehat{z} = 1/\max_j|BCov_n^2(X^{(j)},Y\mid D)|$ is a scale constant. 
We then select the pair of parameters $(\gamma,\lambda_D)$ that minimize the weighted absolute mean difference \citep{shortreed2017outcome}
$$wAMD(\lambda_D, \gamma) = \sum^q_{j=1}|\beta_j|\times\left|\dfrac{\sum^n_{i=1}\widehat{\tau}_i^{\lambda_D,\gamma}X_{i}^{(j)}D_i}{\sum^n_{i=1}\widehat{\tau}_i^{\lambda_D,\gamma}D_i}-\dfrac{\sum^n_{i=1}\widehat{\tau}_i^{\lambda_D,\gamma}X_{i}^{(j)}(1-D_i)}{\sum^n_{i=1}\widehat{\tau}_i^{\lambda_D,\gamma}(1-D_i)}\right|,
$$
where
    $\widehat{\tau}_i^{\lambda_D,\gamma} = {D_i}/{\widehat{e_i}^{\lambda_D,\gamma}} +  (1-D_i)/(1-\widehat{e_i}^{\lambda_D,\gamma})$ and 
 $\widehat{e}_i^{\lambda_D,\gamma}$ is the estimated propensity score with pair of parameters  $(\lambda_D,\gamma)$.
 
 Finally, we use   plug-in estimators $\widehat{e}_i = \widehat{e}(X_i;\widehat{ {\beta}})$ and $\widehat{b}_{d,i} = X_i^\T\widehat{\alpha}^{(d)}$ to construct a doubly robust estimator of $\Delta$ based on equation \eqref{eqn:dr}. These procedures are summarized in Steps 3--6 of Algorithm \ref{alg:cbs}.
 
 %

\begin{algorithm}[t]
  \caption{Causal Ball Screening: A doubly robust estimator of average causal effect with ultra-high dimensional covariates}
  \label{alg:cbs}
\hspace*{0.02in} {\bf Input: }  
 $(X_i,Y_i,D_i)^n_{i=1}$
\hspace*{0.02in} {\bf Output: }
$\widehat{\Delta}$
\begin{algorithmic}[1]
\State For $ j=1, \ldots, p$, calculate $\widehat{\rho}_j = BCov_n^2(X^{(j)},Y\mid D)$.
  \State Select the $q$ variables with the largest $\widehat{\rho}_j$, and denote them as $\K$; without loss of generality, let $\K = \{1,\ldots,q\}$.
  \State For $d=0,1,$ set $\{\widehat{  \alpha }^{(d)}\}^\T =\left(\left(\widehat{{ {\alpha}}}_\K^{(d)}\right)^\T,0^\T\right)$, where $\widehat{{ {\alpha}}}_\K^{(d)}$ is the Lasso estimator obtained via \eqref{eqn:Lasso}.
  \State Set $\widehat{  \beta }^\T =(\widehat{{ {\beta}}}_\K^\T, {0}^\T)$, where $\widehat{ {\beta}}_\K$ is the adaptive Lasso estimator obtained via \eqref{eqn:selection}.
  \State For $i =1,\ldots,n,$ calculate $\widehat{e}_i = \widehat{e}(X_i;\widehat{ {\beta}})$ and $\widehat{b}_{d,i} = {b}_d(X_i;\widehat{\alpha}^{(d)})$.
  \State Plug $\widehat{b}_{d,i}, \widehat{e}_i, d=0,1, i=1,\ldots, n$ into equation \eqref{eqn:dr} to obtain a doubly robust estimator $\widehat{\Delta}$.
\end{algorithmic}
\end{algorithm}

\section{Theoretical Properties}\label{theories}

In this section, we study  theoretical properties of the proposed CBS procedure as outlined in Algorithm \ref{alg:cbs}. We first show the sure independence screening property, which guarantees that 
the set $\K$ selected by the conditional ball covariance screening procedure in Section \ref{sec:condball} includes all the confounders and precision variables with high probability. The following two assumptions are common in the  sure independence screening literature.

\begin{enumerate}
    \item[(A1):] (Minimal strength)  There exist constants $c > 0$ and $0 \leq \kappa<1/2$ such that: $\min\limits_{j\in {\bm{X}^\mC} \cup \bm{X}^\mP}\rho_j\ge2c n^{-\kappa}$, where $\rho_j =BCov^2(X^{(j)},Y\mid D) $;
    \item[(A2):] (Ultra-high dimensional covariates) $\log(p) = o(n^{1-2\kappa})$, where $\kappa$ is defined in (A1).
\end{enumerate}
Condition (A1) specifies the minimum marginal association strength that can be identified by our screening procedure. Condition (A2) allows the dimension of covariates to grow exponentially with the sample size.


 Theorem \ref{cor:fix} shows that we may select the top $q$ variables with the largest $\widehat{\rho}_j$. 
{
\begin{theorem}(Sure independence screening property) \label{cor:fix}
Assume that $W =\{j: \rho_j = 0 \}$ and $|W^c| \leq q,$ where $|\cdot|$ denotes the cardinality. Then under conditions (A1) and (A2), we have 
$
P(\max_{j \in W}\widehat{\rho}_j < \min_{j \in \mathcal{A}}\widehat{\rho}_j) \rightarrow 1
$ and hence $P((\bm{X}^\mC \cup \bm{X}^\mP) \subset \K)\rightarrow 1$ as $n\rightarrow \infty$.
\end{theorem}
}

We then present theoretical guarantees for the variable selection and estimation step. 
 We assume that the data-driven weights $\widehat{w}_j$ and tuning parameters satisfy the following conditions:

\begin{enumerate}
    \item [(B1):] (Convergence inside the target set) For each $j\in \mathcal{A}$, $\widehat{w}_j\stackrel{p}{\longrightarrow}c_j$ , where $c_j$ is a positive constant;
    \item [(B2):] (Uniform convergence to zero outside of the target set) There exists some constant $s>0$ such that for all $j\in \mathcal{A}^c$,  $\widehat{w}_j = O_p(n^{-s})$;
    \item[(B3):] $\lambda_D/\sqrt{n} \rightarrow 0$, $\lambda_Dn^{s-1} \rightarrow \infty $, and for $d=0,1,$ $\lambda_Y^{(d)}/n^{1/2} \rightarrow \infty$, $\lambda_Y^{(d)}/n^{2/3}  \rightarrow 0$; 
    \item [(B4):] $X_\mathcal{A}^\T X_\mathcal{A}/n\overset{\longrightarrow}{}C$, where $C$ is a positive definite matrix;
    \item [(B5):] $\mathcal{A} \subset \mathcal{K}$.
\end{enumerate}
    Conditions (B1) and (B2) are rate conditions on the data-driven weight $\widehat{w}_j$ that are standard in the adaptive Lasso literature \citep[e.g.][]{zou2006adaptive}. Condition (B3) requires that the tuning parameters $\lambda_D, \lambda_Y^{(d)}, d=0,1$ satisfy some rate conditions. Condition (B4) holds as long as  $X_i's$ are i.i.d. with finite second moments. Condition (B5) assumes that the   set $\mathcal{K}$ we select in the screening step includes all the variables in the target adjustment set.  A necessary condition for  (B5) is that the cardinality of the target adjustment set $\mathcal{A}$ is no larger than $q$.  Under this condition, due to Corollary \ref{cor:fix}, Condition (B5) holds 
    with high probability. 

\begin{theorem}\label{thm:selection}
Under Conditions (B1) -- (B5) and Assumptions \ref{assumption:sutva}--\ref{assumption:positivity},  we have:
\begin{enumerate}
    \item [(a).] (Variable selection consistency for the outcome model) Let $\widehat{\mathcal{A}}_{OR} =\{j: \widehat{\alpha}_j^{(d)} \not= 0\; for\; d=0\; or\; 1\}$. Suppose that the outcome regression model satisfies a linear relationship:
\begin{equation}
    \label{eqn:or}
    Y = DX_\mathcal{A}^\T \alpha^{(1)*}_\mathcal{A} + (1-D)X_\mathcal{A}^\T \alpha^{(0)*}_\mathcal{A} + \epsilon, 
\end{equation}
where $\epsilon$ is a random noise with mean $0$ and variance $\sigma^2$.
    Then
$
{\lim}_{n\rightarrow\infty}P(\widehat{\mathcal{A}}_{OR}=\mathcal{A}) =1.
$

\item[(b).] (Variable selection consistency for the propensity score model) Let $\widehat{\mathcal{A}}_{PS} =\{j: \widehat{\beta}_j \not= 0\}$. If the underlying propensity score model $e(X_\mathcal{A})$ is such that 
\begin{equation}
\label{eqn:ps}
e(X_\mathcal{A}) = P(D=1\mid X_\mathcal{A}) = \text{expit}(X_\mathcal{A}^\T \beta_{\mathcal{A}}^*).
\end{equation}
Then
$
{\lim}_{n\rightarrow\infty}P(\widehat{\mathcal{A}}_{PS}=\mathcal{A}) =1.
$
\item [(c).] (Double robustness)  Assume the estimated propensity score model $e(X_i;\widehat{\beta})$ and outcome model $b_d(X_i;\widehat{\alpha}^{(d)})$ converge to some $e^0(X)$ and $b^0_d(X)$ in the sense that 
$$
\dfrac{1}{n}\sum^n_{i=1}\{e(X_i;\widehat{\beta}) - e^0(X_i)\}^2 = o_p(1),\;\dfrac{1}{n}\sum^n_{i=1}\{b_d(X_i;\widehat{\alpha}^{(d)}) - b_d^0(X_i)\}^2 = o_p(1).
$$
Then $
      (\widehat{\Delta} - \Delta)\overset{p}{\longrightarrow}0
    $
if either  \eqref{eqn:or} holds and $\widehat{\mathcal{A}}_{OR}=\mathcal{A}$, or \eqref{eqn:ps} holds and $\widehat{\mathcal{A}}_{OR}=\mathcal{A}$.
\item[(d).] (Oracle asymptotic distribution)
Assume that $\widehat{\mathcal{A}}_{PS} =\widehat{\mathcal{A}}_{OR}= \mathcal{A}$.  Then under models  \eqref{eqn:or} and \eqref{eqn:ps},
\begin{itemize}
    \item[1.] $
    \sqrt{n}(\widehat{\Delta} - \Delta) = \sum^n_{i=1}\{\phi(Y_i,b_1(X_{i,\mathcal{A}}),b_0(X_{i,\mathcal{A}}),e(X_{i,\mathcal{A}}),D_i,\Delta)\}/\sqrt{n} + o_{p}(1);$
    \item[2.] $\sqrt{n}(\widehat{\Delta} - \Delta)V_{\Delta}^{-1/2}\stackrel{d}{\longrightarrow}N(0,1)$;
    \item[3.] $V_{\Delta} - \widehat{V}_{\Delta} = o_{p}(1)$;
    \item[4.] $|{P}\{\Delta\in[\widehat{\Delta}-c_m\widehat{V}_\Delta^{1/2}/\sqrt{n},\widehat{\Delta}+c_m\widehat{V}_\Delta^{1/2}/\sqrt{n}]\}-(1-m)|\rightarrow 0$,
\end{itemize}
\end{enumerate}
\end{theorem}

where $
    \phi(Y,b_1(X),b_0(X),e(X),D,\Delta) =   \dfrac{D \{Y -  b_1(X)\}}{e(X)} -  \dfrac{(1-D)\{ Y - b_0(X)\}}{1-e(X)} + b_1(X) - b_0(X) - \Delta,
$ $\mathbb{E}_n(O) = \dfrac{1}{n}\sum\limits_{i=1}^n O_i,$ $m\in (0,1)$ is the significance level, $c_m = \Phi^{-1} (1-m/2)$,
$V_\Delta = \mathbb{E}\{\phi^2(Y,b_1(X_\mathcal{A}),b_0(X_\mathcal{A}),e(X_\mathcal{A}),D,\Delta)\}$ and $\widehat{V}_\Delta = \mathbb{E}_n\{\phi^2(Y_i,b_1(X_i;\widehat{\alpha}^{(1)}),b_0(X_i;\widehat{\alpha}^{0}),e(X_i;\widehat{\beta}),D_i,\widehat{\Delta})\}$.

\begin{remark}
 Theorem \ref{thm:selection}(a) is parallel to Theorem 1 in \cite{zhao2006model}. Theorem \ref{thm:selection}(b) is parallel to Theorem 4 in \cite{zou2006adaptive} and Theorem 1 in \cite{shortreed2017outcome}. 
\end{remark}

Our results in Theorem \ref{thm:selection} (c) and (d) depend on correct selection of the target  adjustment set $\mathcal{A}.$ 
As such, the resulting uncertainty estimates do not take into account of the uncertainty in the selection of target adjustment set. In other words, as we aim to exclude instruments in our adjustment set, our procedure does not permit uniform valid inference. See \cite{leeb2005model} for discussions of similar phenomena in other contexts involving variable selection.  Alternative procedures that include instruments in the adjustment set are available \citep[e.g.][]{van2014targeted, farrell2015robust,chernozhukov2018double}. In theory, these procedures are uniformly valid and should perform better in the worst-case scenario in which confounders are only weakly related to the outcome. However, in many practical situations, they pay a high price in terms of variance due to inclusion of instrumental variables. 
For example, \cite{brookhart2006variable} show that with small samples,  the inclusion of variables
that are strongly related to the exposure but only weakly related to the outcome can be detrimental
to an estimate in a mean-squared error sense.
We refer readers to \cite{moosavi2021costs} for a recent discussion on the dilemma between uniform validity and efficiency in performing variable selection in causal inference problems. We also illustrate this tradeoff via simulation studies in Section \ref{sim}.

\section{Simulation Studies}\label{sim}


In this section, we evaluate the finite-sample performance of the proposed method. We consider four different combinations of sample size $n$ and covariate dimension $p$: $ (n, p) = (300, 100)$, $(300, 1000)$, $(600, 200)$, $(600, 2000)$. The covariates $X^{(j)}, j=1,\ldots,p$ are independently generated from the uniform distribution on  $(-1,1)$. The binary treatment $D$ is then generated from a Bernoulli distribution with $P(D=1 \mid X ) = \text{expit}(X^\T\beta)$, where $ \beta\in \mathbb{R}^p$ such that $\beta_1 = \beta_2 = 0.2$, $\beta_5 = \beta_6 = 0.3$ and $\beta_j = 0, j\notin \{1,2,5,6\}$. Given $D$ and $X$, the outcome $Y$ is generated from the following model:
$Y = \alpha_0 +  X^\T \alpha +D\Delta+\epsilon$,
where $\epsilon \sim N(0,1)$, $\Delta = 2$, $\alpha_0=0$, $\alpha_j = 2,\; 1\leq j \leq 4$ and $\alpha_j = 0,\; j \geq 5$. Note that in our simulations, both the outcome regression and propensity score models are sparse.

We compare the following methods for estimating the average causal effect:
\begin{enumerate}

\item[(\romannumeral 1)] CBS: The proposed Algorithm \ref{alg:cbs}, where we select the top $ 30$ covariates in Step 2;
\item[(\romannumeral2)] Outcome adaptive Lasso (OAL): We use the {\tt R} code provided in \citet{shortreed2017outcome} to estimate the propensity score. Note that the method by \cite{shortreed2017outcome} cannot directly handle the cases with $p>n.$ For those scenarios, we  first apply a conditional sure-independence screening procedure on $D$ \citep{barut2016conditional} and select the top $ 30$ covariates. We then apply the method of \cite{shortreed2017outcome} on the selected set. We use Lasso to estimate $\widehat{b}_d(X_i)$, and the tuning parameter is selected via $10$-fold cross-validation.

\item[(\romannumeral 3)] Robust Inference \citep[RI,][]{farrell2015robust}: We use Lasso   to fit the propensity score model, and the group-Lasso method implemented in {\tt grpreg} to fit the outcome model. The estimates $\widehat{e}(X_i)$ and $\widehat{b}_d(X_i)$ are then plugged into (\ref{eqn:dr}) to obtain the causal effect estimate. The tuning parameters are selected via $10$-fold cross-validation.

\item[(\romannumeral 4)] Double/Debiased Machine Learning \citep[DBML,][]{chernozhukov2018double}: We implement the cross-fitting estimator with finite-sample adjustment using the median method; here we repeat 10-fold random partitions five times and take their median.



\end{enumerate}

\begin{table}
\thisfloatpagestyle{empty}
\centering
\caption{Simulation results based on $1,000$ Monte Carlo runs. Both the propensity score and outcome regression models are correctly specified. We report  bias $\times 100$, MSE $\times 100$, and the empirical coverage probability for each estimator. The nominal coverage probability is 95\%. Standard errors of bias and MSE are reported in parentheses.   Bold numbers represent the best result in each scenario } 
\bigskip
\scalebox{0.9}{
\begin{tabular}{lllll}
\toprule
(n,p) & Method & $Bias\times100(SE\times100$) & $MSE\times100(SE\times100)$ & $Empirical\; coverage\times 100\%$ \\ \hline
\multirow{4}{*}{$(300,100)$}  
& OAL & $1.3(0.37)$&$\textbf{1.4(0.064)}$&$93.5$ \\
& CBS & $\textbf{0.97(0.38)}$&$1.5(0.066)$&$\textbf{94.3}$\\
& RI  & $1.5(0.76)$&$5.8(0.96)$&$92.5$\\
& DBML*& $8.5(39)$&$1.5\times10^4(2.9\times10^3)$&$98.3$\\ 
& DBML& $2.7\times10^5(2.8\times10^5)$&$7.7\times10^{11}(7.7\times10^{11})$&$98.4$\\\hline
\multirow{4}{*}{$(300,1000)$} 
& OAL & $1.8(0.39)$&$\textbf{1.6(0.075)}$&$\textbf{92.6}$\\
& CBS & $\textbf{1.6(0.40)}$&$1.6(0.076)$&$92.2$\\
& RI  & $19(0.91)$&$12(2.2)$&$51.7$\\
& DBML*&$10\;(37)$&$1.3\times10^4(2.2\times10^3)$&$98.5$\\ 
& DBML&$7.3\times10^2(6.6\times10^2)$&$4.4\times10^6(3.5\times10^6)$&$98.5$ \\

\hline
\multirow{4}{*}{$(600,200)$}  
& OAL     & $0.28(0.26)$&$0.68(0.030)$&$\textbf{94.9}$\\
& CBS     & $\textbf{0.04(0.26)}$&$\textbf{0.68(0.030)}$&$95.6$\\
& RI      & $2.3(0.59)$&$3.6(1.1)$&$91.6$\\
& DBML*    & $16\;(23)$&$5.3\times10^3(1.4\times10^3)$&$98.3$\\
& DBML    & $8.1\times10^2(7.1\times10^2)$&$5.0\times10^6(5.0\times 10^6)$&$98.3$\\
\hline
\multirow{4}{*}{(600,2000)} 
& OAL     & $0.58(0.27)$&$0.71(0.030)$&94.0\\
& CBS     & $\textbf{0.22(0.27)}$&$\textbf{0.71(0.030)}$&$\textbf{94.2}$\\
& RI      & $13(0.51)$&$4.3(0.55)$&$55.9$\\
& DBML*    & $16(1.9\times10^2)$&$2.1\times10^4(1.4\times10^2)$&$98.2$\\
& DBML    & $1.2\times10^2(3.1\times10^2)$&$9.5\times10^5(3.6\times10^5)$&$98.4$\\

\bottomrule
\end{tabular}

}
\begin{tablenotes}
   \item[*] For DBML*, we exclude Monte Carlo runs for which the bias of DBML estimate is greater than 100. For 
   $(n, p) = (300,100), (300,1000), (600,200)$ and $(600,2000),$ we exclude $(35,23,11,26)$ runs out of 1000.
   \end{tablenotes}
\label{Table 1}
\end{table}

Table \ref{Table 1} reports the biases and mean squared errors of various ACE estimators, as well as the empirical coverage  of  $95\%$ Wald-type confidence intervals based on 1,000 Monte Carlo runs. For Table \ref{Table 1}, we fit the correct propensity score and outcome regression models for all four estimating methods. The CBS and OAL perform much better than the RI and DBML methods, suggesting that at least for the simulation settings we consider,  excluding instrumental variables significantly improve the efficiency and reduce the MSE of the resulting doubly robust estimator. We also notice that the performance of the RI method deteriorates quickly with the covariate dimension $p$, while DBML performs the worst under the settings we consider.



\begin{figure}
 \centering
 \includegraphics[width=17cm]{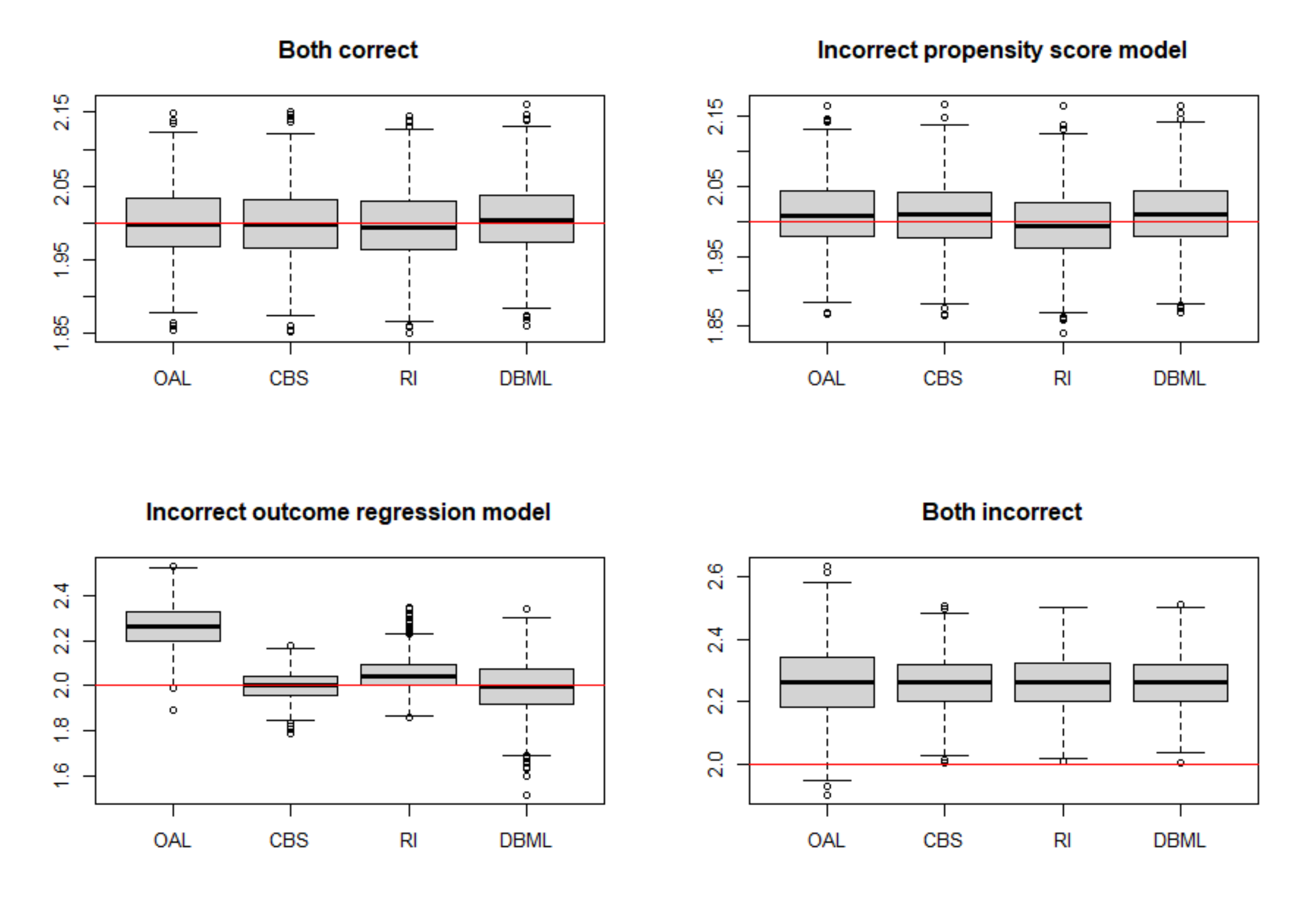}
\caption{Boxplots of causal effect estimates obtained by $\widehat{\Delta}^{OAL}$, $\widehat{\Delta}^{CBS}$, $\widehat{\Delta}^{RI}$ and $\widehat{\Delta}^{DBML}$. The horizontal red lines correspond to the true causal effect $\Delta = 2$. Results are based on 1,000 Monte Carlo runs. This figure appears in color in the electronic version of this article, and any mention of color refers to that version. }
\label{png:CBS}
\end{figure}

We further compare these four estimators in terms of their double robustness. In the following, we consider a low-dimensional setting with $(n,p)=(2000,100)$. The treatment $D$ and outcome $Y$ are generated via the following equations:
\begin{align*}
\text{(Outcome model): \quad \quad \quad }  &Y =  2 (X_1+X_2) + 2(X_3^2+X_4^2)+D\Delta+\epsilon;  \\
\text{(Propensity score model): \quad \quad \quad }  &\text{logit}\{P(D=1\mid X)\} = 0.2(X_1+X_2) + 0.3 (X_5+X_6)+\epsilon.
\end{align*}
To examine double robustness, we  consider settings where the  outcome regression or the propensity score model may be misspecified. In these settings, the analyst assumes that the propensity score model is a logistic regression with predictors $\left\{\left( X^{(j)}\right)^2; j=1,\ldots,p\right\}$, and/or that the outcome model is linear with predictors  $ D $ and $ {\{X^{(j)}}^2; j=1,\ldots, p\}$. 
    Figure \ref{png:CBS} shows boxplots of estimates from the four estimators under   different combinations of correct/incorrect specifications of the outcome regression and propensity score models. One can see that $\widehat{\Delta}^{CBS}$, $\widehat{\Delta}^{RI}$, and $\widehat{\Delta}^{DBML}$ are consistent as long as at least one of the outcome regression or the propensity score model is correctly specified, thus exhibiting double robustness. In contrast,  $\widehat{\Delta}^{OAL}$ is not consistent when the propensity score model is correctly specified but the outcome regression model is not, as it relies on the outcome regression model for confounder selection in the propensity score model.

\section{Real data application}
\label{realdata}

In this session, we analyze  data from the Alzheimer's Disease
Neuroimaging Initiative (ADNI) database (adni.loni.usc.edu). The data use acknowledgement is included in the Data Availability Statement Section. 
We consider the clinical, genetic, and behavioral measures in the ADNI data set. The exposure of interest is the tau protein level in cerebrospinal fluid (CSF) observed at Month 12. Tau is a microtubule-associated protein that promotes microtubule polymerization and stabilization \citep{kametani2018reconsideration}. Studies \citep{iqbal2010tau} have found that tau protein abnormalities initiate the Alzheimer's Disease (AD) cascade and cause neurodegeneration and dementia. Under physiological conditions, tau regulates the assembly and maintenance of the structural stability of microtubules. In a diseased brain, however, tau becomes abnormally hyperphosphorylated, which ultimately causes the microtubules to disassemble, and the free tau molecules aggregate into paired helical filaments \citep{medeiros2011role}. Scientists have found that CSF-tau was markedly increased in Alzheimer's disease \citep{
blennow2015clinical}. In this study, we go beyond association and study whether the CSF-tau protein level affects the severity of the Alzheimer's Disease. 

 We dichotomize the CSF-tau protein level  using the cutoff value $350$ pg/mL  \citep{tapiola2009cerebrospinal}, i.e. $D=1$ if CSF-tau protein level is over $350$ pg/mL, and $D=0$ otherwise. The severity of the AD is measured by the 11-item Alzheimer's Disease Assessment Scale (ADAS-11) cognitive score observed at Month 24, a widely used measure of cognitive behavior  ranging from $0$ to $ 70 $. A higher ADAS-11 score indicates greater severity of AD. 

In our analysis, we adjust for clinical and behavioral covariates, including  baseline age, gender, and education length, as
they are widely considered as the main risk factors for AD \citep[e.g.][]{guerreiro2015age, vina2010women}.
We also consider genetic covariates extracted from whole-genome sequencing data from all of the 22 autosomes.
We provide details of how we preprocess the genetic data in the Supplementary Material. {After pre-processing,  
 $6,087,205$ bi-allelic markers (including SNPs and indels) were retained in the data analysis. }

The data set has $268$ subjects with complete information on  CSF tau protein data, the Month 24 ADAS-11 score and genetic information. Among these subjects, 82 have CSF-tau protein level above the cut-off point. The mean (SD) age in the high/low tau-protein group is   $75.8 (7.28)$ and $75.3 (6.55)$ years old, respectively, and the mean (SD) education length in the high/low tau-protein group is $15.3 (3.05)$ and $15.9 (2.99)$ years, respectively. The two groups are unbalanced in terms of gender: 54.9\% of study participants in the high tau-protein group are female, while 64.0\% of study participants in the low tau-protein group are female. We nevertheless include all three covariates as they were determined {\it a priori.} In the Supplementary Material, we report sensitivity analysis in which we adjust for  more baseline clinical and behavioral covariates.  Analysis results show that adjusting for these additional covariates has  minimal effects on the results we obtained.


Denote  as $ Z$ the   covariates age, gender and education length. We first fit a linear regression $Y\sim  {Z}$ to adjust for these clinical covariates.  We then apply Steps 1-2 of our CBS procedure  using the fitted residuals from the linear regression as the outcome and select the top $30$ genetic covariates. We list the top 10 SNPs in Table \ref{datascreeningresult}. All of them are located on Chromosome 19 and have previously been found to be strongly associated with Alzheimer's. See Table \ref{datascreeningresult2} in the Supplementary Material for a list of references for the SNPs reported in Table \ref{datascreeningresult}. 

\begin{table}
\centering
\caption{The top ten SNPs selected by Steps 1-2 of our CBS procedure} \bigskip
\begin{tabular}{ccccc}
\toprule
Rank & SNP name & Gene & Chromosome number \\ 
\midrule
1 & rs429358    & ApoE    & 19    \\ 
2 & rs56131196  & ApoC1   & 19   \\ 
3 & rs4420638   & ApoC1   & 19     \\ 
4 & rs12721051  & ApoC1   & 19    \\
5 & rs769449    & ApoE    & 19   \\ 
6 & rs10414043  & ApoC1   & 19  \\ 
7 & rs7256200   & ApoC1   & 19   \\ 
8 & rs73052335  & ApoC1   & 19  \\
9 & rs111789331 & ApoC1   & 19   \\ 
10 & rs6857     & NECTIN2 & 19   \\ 
\bottomrule
\end{tabular}
\label{datascreeningresult}
\end{table}

Since some SNPs selected through our first step screening are perfectly correlated, we only keep one among a group of SNPs whose genotypes are identical to each other for the subsequent analysis. We further apply our refined selection procedure in Section \ref{variable selection} on these selected covariates and the covariates age, gender and education length, where the coefficients corresponding to the three clinical and behavioral covariates are not penalized. Finally,  we use the doubly robust estimator \eqref{eqn:dr} to estimate the average causal effect of CSF-tau protein level on the ADAS-11 score. Analysis results suggest that, on average,
being in the high-level CSF-tau group will raise the  ADAS-11 score by $5.96$ ($95\%$ CI = $[4.15, 7.76]$) points. 

In addition to varying the adjusted confounders,  in Section \ref{sec:Sensitivity analysis} of the Supplementary Material, we  also describe sensitivity analysis varying the number of covariates  selected in the first screening step.  Tables \ref{tab:sensitivety analysis: screening} and \ref{tab:sensitive analysis: CI} suggest that   our selection and estimation procedures are relatively robust to the choice of adjusted confounders and  the number of covariates  selected in the   screening step. 






\section{Discussion} 
\label{sec:discussion}

In this paper, we propose a novel selection and estimation procedure for doubly robust causal inference with ultra-high dimensional covariates, called causal ball screening. In comparison to previous approaches that use the same outcome model for propensity score model selection and causal effect estimation, the estimator we propose is doubly robust. Moreover, as  we illustrate in the real data analysis, it can be applied to select variables important for causal inference from millions of baseline covariates.



We have so far considered causal effect estimation using the classical doubly robust estimator by \cite{robins1994estimation}. Our developments  can also be combined with other techniques for causal effect estimation, such as the covariate balancing propensity score  \citep{imai2014covariate} and the subclassification weights \citep{wang2016robust}. This is left as future work.

    \section*{Acknowledgement} 

 Pan was partially supported by the National Natural Science Foundation of China (71991474, 12071494), and the Science and Technology Program of Guangzhou, China  (202002030129, 202102080481).
Kong and Wang were partially supported by the Natural Science and Engineering Research Council of Canada and the CANSSI Collaborative Research Team Grant.  

    \section*{Data Availability Statement}

Data used in preparation of this article were obtained from the Alzheimer's Disease Neuroimaging Initiative (ADNI) database (\url{http://adni.loni.usc.edu/}). As such, the investigators within the ADNI contributed to the design and implementation of ADNI and/or provided data but did not participate in analysis or writing of this report. A complete listing of ADNI investigators can be found at:
\url{http://adni.loni.usc.edu/wp-content/uploads/how_to_apply/ADNI_Acknowledgement_List.pdf}.

\bibliographystyle{biom}
\bibliography{refsg,causal,ref}
\begin{center}
    \section*{Supporting Information}
    \end{center}
    Web Appendices, Tables, Figures and Proofs referenced in Sections \ref{preliminary}, \ref{method}, \ref{theories} and \ref{realdata} are available with this paper at the Biometrics website on Wiley Online Library. The proposed CBS method is available both on Wiley Online Library and on Github: \url{https://github.com/dingketang/ultra-high-DRCI}.




\clearpage

\begin{center}
	
	{\LARGE Supporting Information for ``Ultra-high Dimensional }
	
	{\LARGE  Variable Selection for Doubly Robust Causal Inference'' }
	
	
	
\end{center}
\setcounter{equation}{0}
\setcounter{figure}{0}
\setcounter{table}{0}
\makeatletter
\renewcommand{\theequation}{S\arabic{equation}}
\renewcommand{\thefigure}{S\arabic{figure}}
\renewcommand{\thetable}{S\arabic{table}}
\setcounter{section}{0}

\begin{abstract}
   The supplementary file is organized as follows. Section \ref{sec:Details in target adjustment set} contains background for the target adjustment set discussed in Section \ref{Target set}.  Section \ref{proofproposition} contains the proofs of Propositions \ref{prop:independence}, \ref{prop:unique}--\ref{prop:ignorability}. The proofs of Theorems 1--2 are given in Section \ref{prooftheorem}. Section \ref{additionalrealdata} contains additional information in the real data application including details in genetics data preprocessing, and sensitivity analysis.  
\end{abstract}

\section{Background for the target adjustment set discussed in Section \ref{Target set}}\label{sec:Details in target adjustment set}

We introduce some definitions we shall use in our paper. A DAG is a finite directed graph with no directed cycles. If a directed edge starts from node $X$ and goes to node $Y$, we say $X$ is a parent of $Y$, and Y is a child of $X$. A directed path is a path trace out entirely along with arrows tail-to-head. If there is a directed path from $X$ to $Y$, then $X$ is an ancestor of $Y$, and $Y$ is a descendant of $X$. Next, we define d-separation
\citep{pearl2009causality}.

\begin{definition}(d-separation)
A path is blocked by a set of nodes $Z$ if and only if
\begin{enumerate}
    \item The path contains a chain of nodes $A \rightarrow B \rightarrow C$ or a fork $A \leftarrow B \rightarrow C$ such that the middle node $B$ is in Z (i.e., B is conditioned on), or
    \item  The path contains a collider $A \rightarrow B \leftarrow C$ such that the collision node $B$ is not in $Z$, and no descendant of $B$ is in $Z$.
\end{enumerate}
If $Z$ blocks every path between two nodes $X$ and $Y$, then $X$ and $Y$ are d-separated conditional on $Z$, otherwise $X$ and $Y$ are d-connected conditional on $Z$.
\end{definition}

We introduce some Assumptions under which our definitions of sets have desirable properties.

\begin{assumption}(Causal sufficiency) \label{assumption:sufficiency}
   The causal relationships among $(X, D, Y)$ can be represented by a causal DAG.
\end{assumption}
\begin{assumption}(Temporal ordering)
\label{assumption:order}
    Every $X_i \in X$ is a non-descendant of $D$, which is in turn a non-descendant of $Y$.
\end{assumption}

 \begin{assumption}(Faithfulness)
 \label{assumption:faithfulness}
 The causal DAG encoding the relationships among $(X,D,Y)$ is faithful.
 A distribution P is faithful to a DAG $\mG$ if no conditional independence relations other than the ones entailed by the Markov property are present.
 \end{assumption}

\begin{proposition}\label{prop:unique}
$X^\mC$, $X^\mP$, $X^\mI$, $X^\mN$ are uniquely defined.
\end{proposition}

\begin{proposition} \label{prop:ignorability}
 For any $\mS$ such that $X^\mC \subset X^\mS \subset X^\mC \cup X^\mP \cup X^\mI$, assumptions \ref{assumption:sufficiency} and \ref{assumption:order} imply that $D \ind Y(d)\mid X^\mS, d=0,1.$
\end{proposition}

Proposition \ref{prop:unique} shows unicity of our definition. Proposition \ref{prop:ignorability} shows that our definition of confounders satisfies Property 1 proposed by \cite{vanderweele2013definition}.  Under faithfulness, our definition of confounders also satisfies Property 2A of \cite{vanderweele2013definition}. 
Proposition \ref{prop:ignorability} is an extension of Propositions 1,2,3,4 in \cite{de2011covariate}, the latter imply that $X^\mC$, $X^\mC\cup X^\mI$, $X^\mC\cup X^\mP$, $X^\mC\cup X^\mP\cup X^\mI $ are  valid adjustment sets.





\section{Proofs of Propositions}\label{proofproposition}

\subsection{Proof of Proposition \ref{prop:unique}}
$pa(Y)$ and $pa(D)$ are uniquely defined by definition. For $X^{(i)} \in pa(Y)$, $X^{(i)}$ and $D$ are either d-connected or not given $pa(Y)\setminus X^{(i)}$. For the first case, $X^{(i)}\in X^\mC$, and for the second case, $X^{(i)}\in X^\mP$. Since $pa(D)$ is also uniquely defined, $X^\mI = pa(D)\setminus X^\mC$ is uniquely defined.

\subsection{Proof of Proposition \ref{prop:ignorability} }
We first introduce some graphoid axioms \citep{pearl1985graphoids} we will use later:
\begin{flalign}
    \text{Intersection:}\quad& D \ind Y \mid W,Z; D\ind W\mid Y,Z \Rightarrow D\ind Y,W\mid Z \label{eqn:intersection},\\
        \text{Contraction:}\quad& D \ind Y\mid Z;D\ind W\mid Y,Z \Rightarrow D\ind Y,W\mid Z \label{eqn:contraction},\\
        \text{Weak  union:}\quad&  D \ind X \cup Y \mid Z \Rightarrow D \ind X \mid Z \cup Y, \label{eqn:weakunion}\\
        \text{Decomposition}:\quad&  D \ind X \cup Y \mid Z \Rightarrow D \ind X \mid Z .\label{eqn:decomposition}
\end{flalign}

We first show that any superset of $pa(Y)$ is sufficient to adjust for confounding:
\begin{equation}
\label{eqn:superset}
    D\ind Y(d) \mid \bm{X}^\mathcal{M},
\end{equation}
where $ pa(Y) \subseteq \bm{X}^\mathcal{M}  $.
We show this by contradiction. Assume $D$ and $Y(d)$ are d-connected given $\bm{X}^\mathcal{M}$.
Due to Assumption 2, there is no direct edge between $D$ and $Y(d)$. Furthermore, $D$ and $Y(d)$ are not ancestral to each other due to Assumptions \ref{assumption:order} and \ref{assumption:faithfulness}. Then any path connecting $Y(d)$ and $D$ must be one of the following:
\begin{itemize}
    \item $Y(d)\leftarrow Q  \cdots D $, where $Q$ is a parent of $Y(d).$ Since $Q\in pa(Y)\subset \bm{X}^\mathcal{M}$, this path is blocked by $\bm{X}^\mathcal{M}$;
    \item $Y(d)\rightarrow Q\cdots D $. This is impossible since $X^{{(j)}'}s$ are non-descendants of $Y(d)$.
\end{itemize}

We now show that a precision variable is independent of the treatment conditional on confounders (and other precision variables):
\begin{equation}
    \label{eqn:precision-d-separation}
    D\ind \bm{X}^{\tilde{\mathcal{P}}} \mid \{ pa(Y)\setminus \bm{X}^{\tilde{\mathcal{P}}} \},
\end{equation}
where $\bm{X}^{\tilde{\mathcal{P}}} = pa(Y) \setminus \bm{X}^\mathcal{S}.$

To see this, note that if $ j \in \tilde{\mathcal{P}} \subset \mP$, by the definition of $\mP$ we have $D$ and $X^{(j)}$ are d-separated given $pa(Y)\setminus X^{(j)}$, which implies  $D\ind X^{(j)}\mid \{pa(Y)\setminus X^{(j)}\}$. Without loss of generality, assume ${\tilde{\mathcal{P}}} = \{1,2,3,\ldots,d_0\}$. We then have
\begin{equation*}
    \begin{split}
        &D\ind X^{(1)}\mid [X^{(2)}\cup  \{pa(Y)\setminus X^{(1,2)}\}],\\
        &D\ind X^{(2)}\mid [X^{(1)}\cup  \{pa(Y)\setminus X^{(1,2)}\}].
    \end{split}
\end{equation*}
By the intersection property \eqref{eqn:intersection}, we have $D\ind X^{(1,2)}\mid\{ pa(Y)\setminus X^{(1,2)}\}$. Repeat this process $d_0-1$ time, we then have $D\ind \bm{X}^{\tilde{\mathcal{P}}} \mid\{ pa(Y)\setminus \bm{X}^{\tilde{\mathcal{P}}}\}$.

Combining \eqref{eqn:superset} and \eqref{eqn:precision-d-separation}, by the contraction property \eqref{eqn:contraction}, we can show that adjusting for all the confounders and any precision variables are sufficient to control for confounding:
\begin{equation*}
    \label{eqn:confounders-sufficient}
    D\ind Y(d) \mid \{pa(Y)\setminus \bm{X}^{\tilde{\mathcal{P}}}\}.
\end{equation*}

We now show that an instrument variable set is d-separated from, and hence independent of a precision variable conditional on confounders and other precision variables:
\begin{equation*}
\label{eqn:piindipendence}
\bm{X}^{\tilde{\mI}} \ind X^{(j)} \mid \{pa(Y)\setminus X^{(j)}\},
\end{equation*}
where ${\tilde{\mI}}\subset\mI$, $ j \in \mP$.

We again show by contradiction. Assume there exists $X^{(j)} \in \bm{X}^\mP$ such that $X^{(j)}$ and $\bm{X}^{\tilde{\mI}}$ are d-connected given $pa(Y)\setminus X^{(j)}$. By definition $\bm{X}^\mI\subset pa(D)$, there is a path $D\leftarrow \bm{X}^{\tilde{\mI}}$. Then $D$ and $X^{(j)}$ are d-connected  given $pa(Y)\setminus X^{(j)}$, which is a contradiction to the definition of $\mP$.

We now show that a set of instruments is independent of any subset of precision variables conditional on confounders and other precision variables:
\begin{equation}
\label{eqn:PI d-separation}
    \bm{X}^{\tilde{\mathcal{I}}} \ind \bm{X}^{\tilde{\mathcal{P}}} \mid \{pa(Y)\setminus \bm{X}^{\tilde{\mathcal{P}}}\}, 
\end{equation}
where $\tilde{\mathcal{I}} \subset \mI$.
 Again, without loss of generality, we assume ${\tilde{\mathcal{P}}} = \{1,2,3,\ldots,d_0\}$. We then have:
\begin{equation*}
    \begin{split}
        &\bm{X}^{\tilde{\mathcal{I}}}\ind X^{(1)}\mid [X^{(2)}\cup  \{pa(Y)\setminus X^{(1,2)}\}],\\
        &\bm{X}^{\tilde{\mathcal{I}}}\ind X^{(2)}\mid [X^{(1)}\cup  \{pa(Y)\setminus X^{(1,2)}\}].
    \end{split}
\end{equation*}
By the intersection property \eqref{eqn:intersection}, we have $\bm{X}^{\tilde{\mathcal{I}}}\ind X^{(1,2)}\mid\{ pa(Y)\setminus X^{(1,2)}\}$. Repeat this process $d_0-1$ time, we  have $\bm{X}^{\tilde{\mathcal{I}}}\ind \bm{X}^{\tilde{\mathcal{P}}} \mid\{ pa(Y)\setminus \bm{X}^{\tilde{\mathcal{P}}}\}$.\\

Finally, we show
\begin{equation}
\label{eqn:ignorability, subset of pa(Y) cup pa(D)}
D \ind Y(d) \mid  \bm{X}^{\mS}.
\end{equation}
Using  the same argument as in our proof of \eqref{eqn:superset}, we can show that \begin{equation}
\label{eqn:exposure_precesion_independence}
D\ind \bm{X}^{\tilde{\mP}} \mid [\bm{X}^\mI\cup \{pa(Y)\setminus \bm{X}^{\tilde{\mP}}\}].
\end{equation}
This relationship holds as $pa(D)\subset \bm{X}^\mI\cup \{pa(Y)\setminus \bm{X}^{\tilde{\mP}}\}$, which means $\bm{X}^\mI\cup \{pa(Y)\setminus \bm{X}^{\tilde{\mP}}\}$ is a superset of $pa(D)$. By letting $\tilde{\mI} = \mI$ in \eqref{eqn:PI d-separation}, combining \eqref{eqn:PI d-separation}, \eqref{eqn:exposure_precesion_independence} and the contraction property \eqref{eqn:contraction}, we have
\begin{equation}
\label{eqn:exposure&IV_precesion_independence}    
(\bm{X}^{\mathcal{I}}\cup D)\ind \bm{X}^{\tilde{\mathcal{P}}} \mid\{ pa(Y)\setminus \bm{X}^{\tilde{\mathcal{P}}}\}.
\end{equation}

Combining \eqref{eqn:exposure&IV_precesion_independence} and the decomposition property \eqref{eqn:decomposition}, for $\tilde{\mI}\subset\mI$ we have
\begin{equation*}
(\bm{X}^{\tilde{\mathcal{I}}}\cup D)\ind \bm{X}^{\tilde{\mathcal{P}}} \mid\{ pa(Y)\setminus \bm{X}^{\tilde{\mathcal{P}}}\}.
\end{equation*}
Using the weak union property \eqref{eqn:weakunion}, we have the following result:
$$D\ind \bm{X}^{\tilde{\mathcal{P}}} \mid [\{pa(Y)\setminus \bm{X}^{\tilde{\mathcal{P}}}\} \cup \bm{X}^{\tilde{\mathcal{I}}}].
$$
We set $\tilde{\mathcal{I}} = \mS \cap \mI \subset \mI $. We note that $\{pa(Y)\setminus \bm{X}^{\tilde{\mathcal{P}}}\} \cup (\bm{X}^\mS \cap \bm{X}^\mI) = [pa(Y)\cap \{pa(Y)\cap (\bm{X}^\mS)^c\}^c ] \cup (\bm{X}^\mS \cap \bm{X}^\mI)  = \{pa(Y)\cap \bm{X}^\mS\} \cup (\bm{X}^\mS\cap \bm{X}^\mI) = \bm{X}^\mS $. The last equality holds because $\mS\subset \mP\cup \mI \cup \mC$. So we have
\begin{equation}
\label{eqn:D_precision_independent_givenS}
D\ind \bm{X}^{\tilde{\mathcal{P}}} \mid  \bm{X}^\mS.
\end{equation}

We let $\mathcal{M} = pa(Y)\cup \mS $ in equation \eqref{eqn:superset}. We note that $\bm{X}^\mS \subset \bm{X}^\mathcal{M}$ and $\bm{X}^\mS \cup \bm{X}^{\tilde{\mP}} = \bm{X}^\mathcal{M}$. Combining \eqref{eqn:D_precision_independent_givenS} and \eqref{eqn:superset}, by the contraction property \eqref{eqn:contraction}, we have result \eqref{eqn:ignorability, subset of pa(Y) cup pa(D)}.
$\quad \square$

\subsection{Proof of Proposition \ref{prop:independence}}
Under conditions of Lemma 2.1 in \citep{pan2018ball}  we have
\begin{equation*}
    \begin{split}
        &Bcov(X,Y\mid D) = 0\\
        \Longleftrightarrow & Bcov(X^{(d)},Y^{(d)}) = 0, d = 0,1\\
        \Longleftrightarrow & X^{(d)}\ind Y^{(d)}, d = 0,1 \\ 
        \Longleftrightarrow & X\ind Y\mid D.
    \end{split}
\end{equation*}

\section{Proof of Theorems}\label{prooftheorem}

\subsection{Proof of Theorem \ref{cor:fix}} 


Let $\alpha = BCov^2(X,Y\mid D)$, $\widehat{\alpha} = BCov_n^2(X,Y\mid D)$, $\alpha_1=BCov^2(X^{(1)},Y^{(1)})$ and $\alpha_0=BCov^2(X^{(0)},Y^{(0)})$, where $(X^{(d)}, Y^{(d)})$ follows the same distribution as $(X,Y\mid D=d)$ for $d=0, 1$, respectively. We use $\widehat{\alpha}_1=BCov^2_{n_1}, (X^{(1)},Y^{(1)})$,  $\widehat{\alpha}_0=BCov^2_{n_0} (X^{(0)},Y^{(0)})$ to denote their sample estimators, respectively.
To begin with, there exists a constant $\tilde{c}$ such that 
\begin{equation}
\label{eqn:proof of screening main}
P(|\alpha-\widehat{\alpha}|>cn^{-k})\leq O_p(\exp(-\tilde{c}n^{1-2k})),
\end{equation}
where $c$ and $\kappa$ are defined at condition (A1).

Recall $n_1 = \sum_{i = 1}^n D_i$, $n_0 = n-n_1$, $\omega = P(D=1)$ and $\widehat{\omega} = n_1/n$.

We can write 
\begin{equation*}
\label{eqn:split}
\begin{split}
\alpha - \widehat{\alpha} &= \omega \alpha_1 + (1-\omega)\alpha_0 - \{\widehat{\omega} \widehat{\alpha}_1 + (1-\widehat{\omega})\widehat{\alpha}_0\}\\
&=\omega(\alpha_1-\widehat{\alpha}_1)+(1-\omega)(\alpha_0-\widehat{\alpha}_0)+(\widehat{\alpha}_1-\widehat{\alpha}_0)(\omega-\widehat{\omega}).
\end{split}
\end{equation*}
Since $\alpha_1,\alpha_0,\widehat{\alpha}_1,\widehat{\alpha}_0 \in[0,1]$, $|\alpha - \widehat{\alpha}|\leq \omega|\alpha_1-\widehat{\alpha}_1| + (1-\omega)|\alpha_0-\widehat{\alpha}_0| + |\omega-\widehat{\omega}|$, we have 
\begin{equation}
\label{eqn:3term_thm3}
\begin{split}
P(|\alpha - \widehat{\alpha}|\ge2\epsilon)&\leq P(\omega|\alpha_1-\widehat{\alpha}_1|\ge\omega\epsilon) + P((1-\omega)|\alpha_0-\widehat{\alpha}_0|\ge(1-\omega)\epsilon) + P(|\omega-\widehat{\omega}|\ge\epsilon)\\
&=P(|\alpha_1-\widehat{\alpha}_1|\ge\epsilon) + P(|\alpha_0-\widehat{\alpha}_0|\ge\epsilon) + P(|\omega-\widehat{\omega}|\ge\epsilon).
\end{split}
\end{equation}
We control these terms one by one. For the third term of \eqref{eqn:3term_thm3} 
\begin{equation*}
\omega - \widehat{\omega}  = \frac{1}{n}\sum^{n}_{i=1}(\omega - D_i) = \sum^{n}_{i=1}Z_i,
\end{equation*}
where $Z_i = (\omega-D_i)/n$ are independent zero-mean random variables, and $|Z_i|\leq 1/n=M$, $E(Z_i^2) = \omega(1-\omega)/n^2$. Based on the Bernstein inequality, we have
\begin{equation*}
    P( \omega -  \widehat{\omega}\geq\epsilon) = P(\sum^n_{i=1} Z_i\geq\epsilon)\leq\exp\left(-\frac{\frac{1}{2}\epsilon^2}{\frac{\omega(1-\omega)}{n}+\frac{\epsilon}{3n}}\right).
\end{equation*}

So, we have
\begin{equation}
    \label{enq:thm3_term3}
    P(|\widehat{\omega} - \omega|\geq\epsilon) = P(\sum^n_{i=1} Z_i\geq\epsilon) + P(-\sum^n_{i=1} Z_i\geq\epsilon)\leq2\exp\left(-\frac{\frac{1}{2}\epsilon^2}{\frac{\omega(1-\omega)}{n}+\frac{\epsilon}{3n}}\right).
\end{equation}

\paragraph{}Now we control the first and the second terms of \eqref{eqn:3term_thm3}. Following equation (A.7) from the appendix of \cite{pan2018generic}, there exist two positive constants $c_1$ and $c_2$ such that
\begin{equation*}
\label{eqn:thm3_term12}
    \begin{split}
        &P(|\alpha_1-\widehat{\alpha}_1|\geq\epsilon)\leq 2\exp(-c_1 n_1 \epsilon^2 ),\\
        &P(|\alpha_0-\widehat{\alpha}_0|\geq\epsilon)\leq 2\exp(-c_0 n_0 \epsilon^2 ).
    \end{split}
\end{equation*}

We now show that $\exp(-c_1 n_1 \epsilon^2 ) = O_p(\exp(-c_1 n \omega \epsilon^2/2 ))$. As $\omega = P(D=1) >0$, we have
\begin{equation}
\label{eqn:thm3_term1}
    \begin{split}
    &P\left(\left|\frac{\exp(-c_1 n_1 \epsilon^2 )}{\exp(-c_1 \omega n \epsilon^2/2 )} \right|> 1\right) = P(\frac{ n \omega}{2} -n_1 >0) = P(\omega - \widehat{\omega} > \frac{\omega}{2})\\
    &= P(\sum^{n}_{i=1}Z_i>\frac{\omega}{2})\leq \exp\left( -\frac{\frac{1}{8}\omega^2}{\frac{\omega(1-\omega)}{n}+\frac{\omega}{6n} }    \right)\stackrel{n\rightarrow \infty}{\longrightarrow} 0.
    \end{split}
\end{equation}
Similarly, we have
\begin{equation}
    \label{eqn:thm3_term2}
    \exp(-c_0 n_0 \epsilon^2 ) = O_p(\exp(-c_0 n (1-\omega) \epsilon^2/2 )).
\end{equation}

When $\epsilon < 3 \omega(1-\omega)$, we have 
\begin{equation}
\label{eqn:thm3_rewrite_term3}
    \exp\left(-\frac{\frac{1}{2}\epsilon^2}{\frac{\omega(1-\omega)}{n}+\frac{\epsilon}{3n}}\right) =  \exp\left(-\frac{1}{2\omega(1-\omega)+{2\epsilon}/{3}} n\epsilon^2 \right)\leq \exp(-\tilde{c}_2n\epsilon^2),
\end{equation}
where $\tilde{c}_2 = 1/\{4\omega(1-\omega)\}$. 
Let $\epsilon = cn^{-\kappa}/2$, where $0 < \kappa < 1/2$, $\tilde{c}_1 = c_1\omega/2 $, $ \tilde{c}_0 = c_0  (1-\omega)/2$, combining \eqref{enq:thm3_term3}--\eqref{eqn:thm3_rewrite_term3}, we have
\begin{equation*}
\begin{split}
    P(|\alpha - \widehat{\alpha}|\ge cn^{-\kappa})& \leq O_p(\exp(-\tilde{c}_1cn^{1-2\kappa})) + O_o(\exp(-\tilde{c}_0cn^{1-2\kappa}))\\
    &+ O_p(\exp(-\tilde{c}_2cn^{1-2\kappa})).
\end{split}
\end{equation*}

Let $\tilde{c} = \min(c\tilde{c}_1,c\tilde{c}_0,c\tilde{c}_2)$, we have
\begin{equation*}
\begin{split}
    P(|\alpha - \widehat{\alpha}|\ge cn^{-\kappa})& \leq O_p(\exp(-\tilde{c} n^{1-2\kappa})).
\end{split}
\end{equation*}

Hence, we finish the proof of equation \eqref{eqn:proof of screening main}.
Now let $\rho_j = BCov^2(X^{(j)},Y\mid D)$ and $\widehat{\rho}_j = BCov^2_n(X^{(j)},Y\mid D)$ for $j = 1,2,\ldots,p$. From equation \eqref{eqn:proof of screening main} we know that $P(|\widehat{\rho}_j - \rho_j|>cn^{-\kappa})= O_{p}(\exp(-c_1n^{1-2\kappa}))$.


We then have
\begin{equation*}
    \begin{split}
        P(\max_{j \in W}\widehat{\rho}_j \geq \min_{j \in \mathcal{A}}\widehat{\rho}_j) &= P(\max_{j \in W}\widehat{\rho}_j - \min_{j \in \mathcal{A}}\rho_j  \geq \min_{j \in \mathcal{A}}\widehat{\rho}_j - \min_{j \in \mathcal{A}}\rho_j )\\
        &= P\{(\max_{j \in W}\widehat{\rho}_j -0) - (\min_{j \in \mathcal{A}}\widehat{\rho}_j - \min_{j \in \mathcal{A}}\rho_j)  \geq  \min_{j \in \mathcal{A}}\rho_j\}\\
        &\leq P\{(\max_{j \in W}\widehat{\rho}_j -0) - (\min_{j \in \mathcal{A}}\widehat{\rho}_j - \min_{j \in \mathcal{A}}\rho_j)  \geq  2cn^{-\kappa}\}\\
        &\leq P\{(\max_{j \in W}\widehat{\rho}_j -0)> cn^{-\kappa} \} +P\{-\min_{j \in \mathcal{A}}\widehat{\rho}_j + \min_{j \in \mathcal{A}}\rho_j  \geq cn^{-\kappa}\}\\
        &\leq P\{\max_{j \in W}|\widehat{\rho}_j -\rho_j|)> cn^{-\kappa}\} + P\{\max_{j \in A}|\widehat{\rho}_j -\rho_j|)\geq cn^{-\kappa}\} \\
        &\leq O(p\exp(-\tilde{c}n^{1-2\kappa})).
    \end{split}
\end{equation*}
The first inequality holds given condition (A1), and the last inequality holds given equation  \eqref{eqn:proof of screening main}. Given above results, we have 
$$
P(\max_{j \in W}\widehat{\rho}_j < \min_{j \in \mathcal{A}}\widehat{\rho}_j) \geq 1- O(p\exp(-\tilde{c}n^{1-2\kappa})).
$$

As a result, we have 
$$
P(\mathcal{A}\subset\widehat{\mathcal{A}})\geq P(\max_{j \in W}\widehat{\rho}_j < \min_{j \in \mathcal{A}}\widehat{\rho}_j) \geq 1- O(p\exp(-\tilde{c}n^{1-2\kappa})).
$$
\subsection{Proof of Theorem \ref{thm:selection}} 
\paragraph{} Claims (a) and (b) of Theorem  \ref{thm:selection} are direct applications of the consistency of Lasso \citep[][Theorem 1]{zhao2006model} and adaptive Lasso for logistic regression \citep[][Theorem 1]{shortreed2017outcome} respectively. We now show claims (c) and (d).

We first show the following results:
\begin{enumerate}
    \item[(i).] If \eqref{eqn:or} holds and $\widehat{\mathcal{A}}_{OR}=\mathcal{A}, $ then $\frac{1}{n}\sum^n_{i=1}\{b_d(X_i,\widehat{\alpha}^{(d)}) - b_d^0(X_i,\alpha^{(d)})\}^2 = O_p(1/n^{2/3})$;
    \item[(ii).] If \eqref{eqn:ps} holds and $\widehat{\mathcal{A}}_{PS}=\mathcal{A}, $ then $\frac{1}{n}\sum^n_{i=1}\{e(X_i,\widehat{\beta}) - e^0(X_i,\beta)\}^2 = O_p(1/n)$.
\end{enumerate}
\paragraph{Proof of (i):}
\begin{equation*}
\begin{split}
    \frac{1}{n}\sum^n_{i=1}\{b_d(X_i,\widehat{\alpha}^{(d)}) - b_d(X_i,\alpha^{(d)*})\}^2 &= \frac{1}{n}\sum^n_{i=1}\{X_{i,\mathcal{A}}^\T(\widehat{\alpha}^{(d)}_\mathcal{A} - {\alpha}^{(d)*}_\mathcal{A})\}^2\\
    &= (\widehat{\alpha}^{(d)}_\mathcal{A} - {\alpha}^{(d)*}_\mathcal{A})^\T(\frac{1}{n} X_\mathcal{A}^\T X_\mathcal{A})(\widehat{\alpha}^{(d)}_\mathcal{A} - {\alpha}^{(d)*}_\mathcal{A})\\
    &\leq O_p(\frac{1}{n^{2/3}}).
\end{split}
\end{equation*}
The last inequality holds because $(\frac{1}{n} X_\mathcal{A}^\T X_\mathcal{A})\overset{p}{\longrightarrow}C $ given condition (B4), and $(\widehat{\alpha}^{(d)}_\mathcal{A} - {\alpha}^{(d)*}_\mathcal{A})\leq O_p(\lambda^{(d)}_Y/n)$ given lemma 3 in \cite{zou2006adaptive}. 

\paragraph{Proof of (\rmnum{2}):}
\begin{equation*}
\begin{split}
    \frac{1}{n}\sum^n_{i=1}\{e(X_i,\widehat{\beta}) - e(X_i)\}^2 &=  \frac{1}{n}\sum^n_{i=1}\{\text{expit}(X_{i,\mathcal{A}}^\T\widehat{\beta}_\mathcal{A}) - \text{expit}(X_{i,\mathcal{A}}^\T\beta^*_\mathcal{A})\}^2\\
    &\leq \frac{1}{n}\sum^n_{i=1}\{X_{i,\mathcal{A}}^\T\widehat{\beta}_\mathcal{A} - X_{i,\mathcal{A}}^\T\beta^*_\mathcal{A}\}^2\\
    &= (\widehat{\beta}_\mathcal{A} - {\beta}^{*}_\mathcal{A})^\T(\frac{1}{n} X_\mathcal{A}^\T X_\mathcal{A})(\widehat{\beta}_\mathcal{A} - {\beta}^{*}_\mathcal{A})\\
    &\leq O_p(1/n).
\end{split}
\end{equation*}

The last inequality holds because $(\frac{1}{n} X_\mathcal{A}^\T X_\mathcal{A}) \overset{p}{\longrightarrow}C  $ given condition (B4), and $(\widehat{\beta}_\mathcal{A} - {\beta}^*_\mathcal{A})\leq O_p(1/\sqrt{n})$ given   \citet[][Theorem 2(b)]{shortreed2017outcome}.

\paragraph{Proof of Claim (c):}

We have the following decomposition:

\begin{equation*}
    \begin{split}
        \widehat{\Delta} - \Delta  &=  \frac{1}{n}\sum^n_{i=1}\phi(Y_i,b^0_1(X_i),b^0_0(X_i),e^0(X_i),D_i,\Delta)+R_1+R_2+R_3+R_4,
    \end{split}
\end{equation*}

where
\begin{equation*}
    \begin{split}
        R_1 &= \frac{1}{n}\sum^n_{i=1}\{b_1(X_i,\widehat{\alpha}^{(1)})-b_1^0(X_i)\} - \frac{1}{n}\sum^n_{i=1}\{b_0(X_i,\widehat{\alpha}^{(0)})-b_0^0(X_i)\},\\
        R_2 &= \frac{1}{n}\sum^n_{i=1}\frac{D_iY_i}{{e(X_i,\widehat{\beta})e^0(X_i)}}\{e^0(X_i) -e(X_i,\widehat{\beta})\}\\ &+\frac{1}{n}\sum^n_{i=1}\frac{(1-D_i)Y_i}{\{1-e(X_i,\widehat{\beta})\}\{1-e^0(X_i)\}}\{e^0(X_i) - e(X_i,\widehat{\beta})\},\\
        R_3&= -\frac{1}{n}\sum^n_{i=1} \frac{D_i}{e(X_i,\widehat{\beta})}\{b_1(X_i,\widehat{\alpha}^{(1)})-b_1^0(X_i)\}+\frac{1}{n}\sum^n_{i=1} \frac{1-D_i}{1-e(X_i,\widehat{\beta})}\{b_0(X_i,\widehat{\alpha}^{(0)})-b_0^0(X_i)\},\\
        R_4 &= \frac{1}{n}\sum^n_{i=1}\frac{D_ib_1^0(X_i)}{e(X_i,\widehat{\beta})e^0(X_i)}\{e(X_i,\widehat{\beta})-e^0(X_i)\} \\
        &+ \frac{1}{n}\sum^n_{i=1}\frac{(1-D_i)b_0^0(X_i)}{\{1-e(X_i,\widehat{\beta})\}\{1-e^0(X_i)\}}\{e(X_i,\widehat{\beta})-e^0(X_i)\}.
    \end{split}
\end{equation*}

Under Assumption \ref{assumption:positivity}, the Cauchy-Schwarz inequality and the consistency condition of Theorem \ref{variable selection} (c),
$R_1$, $R_2$, $R_3$, $R_4$ are $o_p(1)$.

$$
\frac{1}{n}\sum^n_{i=1}\phi(Y_i,b^0_1(X_i),b^0_0(X_i),e^0(X_i),D_i,\Delta)\overset{p}{\longrightarrow} \Delta
$$
if $b^0_d(X_i) = b_d(X_i,\alpha^{(d)^*})$ or $e^0(X_i) = e(X_i,\beta^*)$, which holds by claim (\rmnum{1}) and (\rmnum{2}).

\paragraph{Proof of Claim (d):}

We have the following decomposition:
\begin{equation*}
    \begin{split}
        \sqrt{n}(\widehat{\Delta} - \Delta) - \sum^n_{i=1}\{\phi(Y_i,b_1(X_{i,\mathcal{A}}),b_0(X_{i,\mathcal{A}}),e(X_{i,\mathcal{A}}),D_i,\Delta)\}/\sqrt{n}= R_1 + R_{21} + R_{22},
    \end{split}
\end{equation*}
where
\begin{equation*}
    \begin{split}
    R_1 &= \frac{1}{\sqrt{n}}\sum^n_{i=1}D_i\{Y_i - b_1(X_i,\alpha^{(1)*})\} \{\frac{1}{e(X_i,\widehat{\beta})}-\frac{1}{e(X_i,\beta^*)}\}\\
    &- \frac{1}{\sqrt{n}}\sum^n_{i=1}(1-D_i)\{Y_i - b_0(X_i,\alpha^{(0)*})\} \{\frac{1}{1-e(X_i,\widehat{\beta})}-\frac{1}{1-e(X_i,\beta^*)}\},\\
    R_{21} &= \frac{1}{\sqrt{n}}\sum^n_{i=1}\left[\{b_1(X_i,\widehat{\alpha}^{(1)}) - b_1(X_i,\alpha^{(1)*})\}\frac{e(X_i,\beta^*)-D_i}{e(X_i,\beta^*)}\right]\\
    &+ \frac{1}{\sqrt{n}}\sum^n_{i=1}\left[\{b_0(X_i,\widehat{\alpha}^{(1)}) - b_0(X_i,\alpha^{(1)*})\}\frac{e(X_i,\beta^*)-D_i}{e(X_i,\beta^*)}\right],\\
    R_{22} &= \frac{1}{\sqrt{n}}\sum^n_{i=1}\left[\{b_1(X_i,\widehat{\alpha}^{(1)}) - b_1(X_i,\alpha^{(1)*})\}\{e(X_i,\widehat{\beta})-e(X_i,\beta^*)\}\frac{D_i}{e(X_i,\widehat{\beta})e(X_i,\beta^*)}\right]\\
    +&\frac{1}{\sqrt{n}}\sum^n_{i=1}\left[\{b_0(X_i,\widehat{\alpha}^{(1)}) - b_0(X_i,\alpha^{(1)*})\}\{e(X_i,\widehat{\beta})-e(X_i,\beta^*)\}\frac{1-D_i}{(1-e(X_i,\widehat{\beta}))(1-e(X_i,\beta^*))}\right].
    \end{split}
\end{equation*}
Due to symmetry, we only show that  the first term of $R_{1},\; R_{21},\; R_{22}$ are $o_p(1)$ using (\rmnum{1}) and (\rmnum2).

For $R_1$, let the first and second term of $R_1$ be $R_{11}$ and $R_{12}$, respectively. We have 

\begin{equation*}
    \begin{split}
    \mathbb{E}(R_{11}^2\mid \{X_i,D_i\}^n_{i=1}) =& \frac{1}{n}\sum^n_{i=1}\frac{D_i\sigma^2}{e^2(X_i,\beta^*)e^2(X_i,\widehat{\beta})}\{e(X_i,\widehat{\beta})-e(X_i,\widehat{\beta}^*)\}^2\\
    \leq& \frac{C}{n}\sum^n_{i=1}\{e(X_i,\widehat{\beta})-e(X_i,\widehat{\beta}^*)\}^2\\
    \leq& o_p(1).
    \end{split}
\end{equation*}
Because $\mathbb{E}(R_{11}\mid \{X_i,D_i\}^n_{i=1}) = 0$, we have $R_{11} = o_p(1)$. 

Similarly, let the first and second term of $R_{21}$ be $R_{211}$ and $R_{212}$, respectively. We have

\begin{equation*}
\begin{split}
 \mathbb{E}(R_{211}^2\mid \{X_i,D_i\}^n_{i=1}) &= \frac{1}{n}\sum^n_{i=1} \left[\{b_1(X_i,\widehat{\alpha}^{(1)}) - b_1(X_i,\alpha^{(1)*})\}^2\left\{\frac{e(X_i,\beta^*)-D_i}{e(X_i,\beta^*)}\right\}^2\right]\\
 &\leq\frac{C}{n}\sum^n_{i=1} \{b_1(X_i,\widehat{\alpha}^{(1)}) - b_1(X_i,\alpha^{(1)*})\}^2\\
 &\leq o_p(1).
\end{split}
\end{equation*}

By the same argument as $R_{11}$, we can show that $R_{211}$ is $o_p(1)$.

For $R_{22}$, by the Cauchy-Schwarz inequality,
\begin{equation*}
    \begin{split}
        &\frac{1}{\sqrt{n}}\sum^n_{i=1}\left[\{b_1(X_i,\widehat{\alpha}^{(1)}) - b_1(X_i,\alpha^{(1)*})\}\{e(X_i,\widehat{\beta})-e(X_i,\beta^*)\}\frac{D_i}{e(X_i,\widehat{\beta})e(X_i,\beta^*)}\right]\\
        \leq&\frac{1}{\sqrt{n}}\max \{e(X_i,\widehat{\beta})e(X_i,\beta^*)\}\sqrt{\sum^n_{i=1}\{b_1(X_i,\widehat{\alpha}^{(1)}) - b_1(X_i,\alpha^{(1)*})\}^2\sum^n_{i=1}\{e(X_i,\widehat{\beta})-e(X_i,\beta^*)\}^2}\\
        \leq& O_p(\frac{1}{\sqrt{n}})O_p(1)O_p(n^{1/6})\\
        \leq&o_p(1).
    \end{split}
\end{equation*}
Hence, we have finished the proof of Theorem  \ref{thm:selection} (d.1). For  Theorem  \ref{thm:selection} (d.3), the proof is very similar to the ones in the proof of Theorem 3.3 in \cite{farrell2015robust}.  Proofs for the remaining  parts of Theorem 2 are straightforward and hence omitted.

\section{Additional real data results} \label{additionalrealdata}

\subsection{Genetics data preprocessing}

For these genetic data, we applied the following preprocessing technique. The first line quality control steps include (i) call rate check per subject and per Single Nucleotide Polymorphism (SNP) marker, 
(ii) gender check, (iii) sibling pair identification, (iv) the Hardy-Weinberg equilibrium test, (v) marker removal by the minor allele frequency, and (vi) population stratification. The second line preprocessing steps include removal of SNPs with (i) more than 5$\%$ missing values, (ii) minor allele frequency (MAF) smaller than 10$\%$, and (iii) Hardy-Weinberg equilibrium p-value $< 10^{-6}$. $503,892$ SNPs obtained from 22 chromosomes were included in for further processing. MACH-Admix software (http://www.unc.edu/~yunmli/MaCH-Admix/) \citep{LiuLi2013} is applied  to perform genotype imputation, using 1000G Phase I Integrated Release Version 3 haplotypes (http://www.1000genomes.org) \citep{GPC1000} as a reference panel. Quality control was also conducted after imputation, excluding markers with (i) low imputation accuracy (based on imputation output $R^2$), (ii) Hardy-Weinberg equilibrium p-value $10^{-6}$, and (iii) minor allele frequency (MAF) $<5\%$.

\subsection{Related literature for the top 10 SNPs listed in Table \ref{datascreeningresult}}

In Table \ref{datascreeningresult2}, we include references for the top 10 listed in Table \ref{datascreeningresult}.

\begin{table}
\centering
\caption{The top ten SNPs selected by our CBS procedure} \bigskip
\begin{tabular}{ccccc}
\toprule
Rank & SNP name & Gene & Chromosome number & Related references \\ 
\midrule
1 & rs429358    & ApoE    & 19    & \cite{cramer2012apoe} \\ 
2 & rs56131196  & ApoC1   & 19    & \cite{guerreiro2012genetic} \\ 
3 & rs4420638   & ApoC1   & 19    & \cite{guerreiro2012genetic}  \\ 
4 & rs12721051  & ApoC1   & 19    & \cite{gao2016shared}\\
5 & rs769449    & ApoE    & 19    & \cite{cruchaga2013gwas}\\ 
6 & rs10414043  & ApoC1   & 19    & \cite{zhou2014association}\\ 
7 & rs7256200   & ApoC1   & 19    & \cite{takei2009genetic}\\ 
8 & rs73052335  & ApoC1   & 19    & \cite{zhou2014association}\\
9 & rs111789331 & ApoC1   & 19    & \cite{rajabli2018ancestral} \\ 
10 & rs6857     & NECTIN2 & 19    & \cite{kamboh2012genome}\\ 
\bottomrule
\end{tabular}
\label{datascreeningresult2}
\end{table}

\subsection{Sensitivity analysis} \label{sec:Sensitivity analysis}
We perform sensitivity analysis by varying the number of observed confounders adjusted and the number of covariates kept in the first screening step. In the analyses reported in Section \ref{realdata} of the main paper, we adjust for the observed confounders baseline age, gender, and education length as they are the main risk factors for AD \citep{guerreiro2015age, vina2010women}. In the dataset, there are other covariates available including handedness, retirement status, and martial status. In the sensitivity analysis, we  adjust for these additional covariates in addition to the age, gender, and education length. We also perform another set of sensitivity analyses by setting a different threshold for the cardinality of set $\K$ selected by the first screening step. In addition to the number $30$ used in the main paper, we also consider $q=40$ ,$50$, or $\lfloor n/\log(n)\rfloor=47$, the last of which was suggested by \citet{fan2008sure}. The results are included in Table \ref{tab:sensitivety analysis: screening}. The estimated average causal effects are similar in all settings.

\begin{table}
\centering
\caption{The top ten SNPs selected by our CBS procedure with additional clinical confounders} \bigskip
\scalebox{0.8}{
\begin{tabular}{cccccc}
\toprule
Rank & SNP name & Gene & Chromosome number & Related references & Rank in the old procedure \\ 
\midrule
1 & rs429358    & ApoE    & 19    & \cite{cramer2012apoe}      &1  \\ 
2 & rs12721051  & ApoC1   & 19    & \cite{gao2016shared}       &4\\
3 & rs56131196  & ApoC1   & 19    & \cite{guerreiro2012genetic}&2 \\ 
4 & rs4420638   & ApoC1   & 19    & \cite{guerreiro2012genetic}&3  \\ 
5 & rs769449    & ApoE    & 19    & \cite{cruchaga2013gwas}    &5 \\ 
6 & rs10414043  & ApoC1   & 19    & \cite{zhou2014association} &6 \\ 
7 & rs7256200   & ApoC1   & 19    & \cite{takei2009genetic}    &7\\ 
8 & rs73052335  & ApoC1   & 19    & \cite{zhou2014association} &8 \\
9 & rs111789331 & ApoC1   & 19    & \cite{rajabli2018ancestral}&9 \\ 
10 & rs6857     & NECTIN2 & 19    & \cite{kamboh2012genome}    &10\\ 
\bottomrule
\end{tabular}
}
\label{tab:sensitivety analysis: screening}
\end{table}


Table \ref{tab:sensitivety analysis: screening} shows the top 10
SNPs from the screening step with additional clinical confounders. One can see immediately that Table \ref{tab:sensitivety analysis: screening}  is quite similar to Table \ref{datascreeningresult2}. We get the same SNPs with different orders. Table \ref{tab:sensitive analysis: CI} provides point estimates and confidence intervals with different  clinical confounders and thresholds for the cardinality of $|\K|$. 
These estimates are close to each other, suggesting that our method is relatively robust to the clinical/behavioral confounders we adjust for, and the number of covariates we keep in the screening step.

\begin{table}[h]
\centering
\caption{Point estimates and 95\% confidence interval with different clinical/behavioral covariates  and cardinality of $\K$ }\bigskip
\begin{tabular}{llll}
\toprule
Clinical confounders & Threshold $|\K|$ & Point estimate & 95\% confidence interval \\ \midrule
\multirow{4}{*}{\begin{tabular}[c]{@{}l@{}}age,gender,\\ education length\end{tabular}}                             
&30           & 5.96&[4.15, 7.76]  \\\cline{2-4}
&40           & 5.95&[4.25, 7.64]   \\\cline{2-4}
&50           & 6.08&[4.25, 7.91]\\\cline{2-4} 
&47$(\lfloor n/\log(n)\rfloor)$&6.08&[4.30, 7.90]   \\\midrule
\multirow{4}{*}{\begin{tabular}[c]{@{}l@{}}age,gender, education\\ length, handedness,\\ retirement status,\\  marital status\end{tabular}} 
&30            & 6.68 & [4.96, 8.40] \\\cline{2-4} 
&40            & 6.76 & [5.09, 8.43]  \\\cline{2-4} 
&50            & 6.75 & [5.06, 8.44]   \\\cline{2-4}             &47$(\lfloor n/\log(n)\rfloor)$             &             6.76&[5.05, 8.48]  \\
\bottomrule
\end{tabular}
\label{tab:sensitive analysis: CI}
\end{table}


\end{document}